\documentstyle[12pt,aaspp4]{article}

\newcommand{\citep}{\cite}
\newcommand{\citet}{\cite}

\lefthead{Klessen, R.~S.}  \righthead{PDF's of Supersonic Turbulence
in Self-Gravitating Media}

\begin{document}

\title{One-Point Probability Distribution Functions of Supersonic
Turbulent Flows in Self-Gravitating Media}
\author{Ralf S.~Klessen}
\affil{Sterrewacht Leiden, Postbus 9513, 2300-RA Leiden, The
Netherlands\\
Max-Planck-Institut f{\"u}r Astrophysik, K{\"o}nigstuhl 17, 69117
Heidelberg, Germany
}

\begin{center}
{\em (accepted for publication in The Astrophysical Journal)}\hfill
\end{center}

\begin{abstract}
Turbulence is essential for understanding the structure and dynamics
of molecular clouds and star-forming regions. There is a need for
adequate tools to describe and characterize the properties of
turbulent flows. One-point probability distribution functions (pdf's)
of dynamical variables have been suggested as appropriate statistical
measures and applied to several observed molecular clouds.  However,
the interpretation of these data requires comparison with numerical
simulations.  To address this issue, SPH simulations of driven and
decaying, supersonic, turbulent flows with and without self-gravity
are presented. In addition, random Gaussian velocity fields are
analyzed to estimate the influence of variance effects.  To
characterize the flow properties, the pdf's of the density, of the
line-of-sight velocity centroids, and of the line centroid increments
are studied.  This is supplemented by a discussion of the dispersion
and the kurtosis of the increment pdf's, as well as the spatial
distribution of velocity increments for small spatial lags. From the
comparison between different models of interstellar turbulence, it
follows that the inclusion of self-gravity leads to better agreement
with the observed pdf's in molecular clouds.  The increment pdf's for
small spatial lags become exponential for all considered
velocities. However, all the processes considered here lead to
non-Gaussian signatures, differences are only gradual, and the
analyzed pdf's are in addition projection dependent. It appears
therefore very difficult to distinguish between different physical
processes on the basis of pdf's only, which limits their applicability
for adequately characterizing interstellar turbulence.
\end{abstract}


\section{Introduction}
\label{sec:intro}
Turbulence is an important ingredient for understanding the properties
and characteristics of molecular clouds and star-forming
regions. Turbulent gas motions are highly supersonic as indicated by
the superthermal line widths ubiquitously observed throughout
molecular clouds (Williams, Blitz \& McKee 2000). These motions carry
enough energy to halt global collapse and act as stabilizing agent for
the entire cloud. However, it can be shown that interstellar
turbulence decays quite rapidly on time scales of the order of the
free-fall time of the system (Mac~Low et al.~1998, Stone, Ostriker \&
Gammie 1998, Padoan \& Nordlund 1999). To explain the observed long
life times, turbulence in molecular clouds must be constantly driven
(Gammie \& Ostriker 1996, Mac~Low 1999). The interplay between
self-gravity on the one hand (leading to local collapse and star
formation) and turbulent gas motion on the other hand (trying to
prevent this process) plays a key role in determining the structure of
molecular clouds.  Altogether, understanding the characteristics of
compressible, supersonic, and constantly replenished turbulence in
self-gravitating media is an important ingredient for an adequate
description of molecular clouds dynamics. And vice versa, from
analyzing the spatial and dynamical structure of molecular clouds we
can gain insight into the phenomenon of turbulence (for an overview
over interstellar turbulence see Franco \& Carraminana 1999).

Unfortunately, a complete and comprehensive theory of turbulence does
not exist. Due to the enormous complexity of the problem, progress has
been slow since Kolmogorov's pioneering work in 1941, where he derived
simple scaling laws for incompressible, stationary, and homogeneous
turbulence by postulating a self-similar energy cascade downwards from
the driving scale to the dissipation range. Most effort has since been
put in finding an adequate closure procedure, i.e.~in finding a way to
express the highest-order correlation in the hierarchy of equations
governing turbulent motion (for an excellent overview see Lesieur
1997; also Boratav, Eden \& Erzan 1997). However, a satisfying
description of turbulence has yet to be found.

Correlation and distribution functions of dynamical variables are
frequently deployed for characterizing the kinematical properties of
turbulent molecular clouds.  Besides using 2-point statistics
(e.g.~Scalo 1984, Kleiner \& Dickman 1987, Kitamura et al.~1993,
Miesch \& Bally 1994, LaRosa, Shore \& Magnani 1999), many studies
have hereby concentrated on 1-point statistics, namely on analyzing
the probability distribution function (pdf) of the (column) density
and of dynamical observables, e.g.~of the centroid velocities of
molecular lines and their increments.  The density pdf has been used
to characterize numerical simulations of the interstellar medium by
V{\'a}zquez-Semadeni (1994), Padoan, Nordlund, \& Jones (1997),
Passot, \& V{\'a}zquez-Semadeni (1998) and Scalo et
al.~(1998). Velocity pdf's for several star-forming molecular clouds
have been determined by Miesch \& Scalo (1995) and Miesch, Scalo \&
Bally (1998). Lis et al.~(1996, 1998) analyzed snapshots of a
numerical simulation of mildly supersonic, decaying turbulence
(without self-gravity) by Porter, Pouquet, \& Woodward (1994) and
applied the method to observations of the $\rho$-Ophiuchus
cloud. Altogether, the observed pdf's exhibit strong non-Gaussian
features, they are often nearly exponential with possible evidence for
power-law tails in the outer parts. This disagrees with the nearly
Gaussian behavior typically found in experimental measurements and
numerical models of incompressible turbulence. The observed centroid
velocity {\em increment} pdf's are more strongly peaked and show
stronger deviations from Gaussianity than numerical models of
incompressible turbulence predict.  Furthermore, the spatial
distribution of the largest centroid velocity differences (determining
the tail of the distribution) appears `spotty' across the face of the
clouds; there is no convincing evidence for filamentary
structure. Miesch et al.~(1998) conclude that turbulence in molecular
clouds involves physical processes that are not adequately described
by incompressible turbulence or mildly supersonic decay simulations
(see also Mac~Low \& Ossenkopf 2000).

It is the principal goal of this paper to extend previous
determinations of pdf's from numerical models into a regime more
applicable for interstellar turbulence by (1) by calculating fully
supersonic flows, (2) by including self-gravity, and (3) by
incorporating a (simple analytic) description of turbulent energy
input.  For comparison with molecular cloud observations, I discuss
the dynamical properties of decaying and stationary (i.e.~driven),
supersonic, isotropic turbulence in self-gravitating isothermal
gaseous media. The pdf's for the density, for the line centroid
velocity and for their increments are derived as function of time and
evolutionary state of the turbulent model.

The structure of this paper is as follows: Section~\ref{sec:remarks}
introduces and defines the statistical tools applied in the current
study. It is followed in Sec.~\ref{sec:model} by a description of the
numerical scheme used to compute the time evolution of the turbulent
flows. Sec.~\ref{sec:init} shows that already simple variance effects
in random Gaussian fields are able to introduce strong {\em
non}-Gaussian distortions to the pdf's which makes a clear-cut
interpretation difficult. Section \ref{sec:decay-non-grav} contains
the analysis of decaying, initially highly supersonic turbulence
without self-gravity. This effect is then added to the simulations
presented in Sec.~\ref{sec:decay-with-gravity}. The model most
relevant for molecular cloud dynamics is discussed in
Sec.~\ref{sec:driven-with-gravity}. It includes a simple driving term
to replenish the turbulent cascade. Finally, in Sec.~\ref{sec:summary}
all results are summarized.

\section{PDF's and Their Interpretation}
\label{sec:remarks}
\subsection{Turbulence and PDF's}
\label{subsec:turbulence}

The Kolmogorov (1941) approach to incompressible turbulence is a
purely phenomenological one and assumes the existence of a stationary
turbulent cascade. Energy is injected into the system at large scales
and cascades down in a self-similar way. At the smallest scales it
gets converted into heat by molecular viscosity. The flow at large
scales is essentially inviscid, hence for small wave numbers the
equation of motion is dominated by the advection term.  If the
stationary state of fully developed turbulence results from random
external forcing then one na\"{\i}vely expects the velocity
distribution in the fluid to be Gaussian on time scales larger than
the correlation time of the forcing, irrespectively of the statistics
of the forcing term which follows from the central limit theorem.
However, the situation is more complex (e.g.\ Frisch 1995, Lesieur
1997). One of the most striking (and least understood) features of
turbulence is its intermittent spatial and temporal behavior. The
structures that arise in a turbulent flow manifest themselves as high
peaks at random places and at random times. This is reflected in the
pdf's of dynamical variables or passively advected scalars.  They are
sensitive measures of deviations from Gaussian statistics. Rare strong
fluctuations are responsible for extended tails, whereas the much
larger regions of low intensity contribute to the peak of the pdf near
zero (for an analytical approach see e.g. Forster, Nelson \& Stephens
1977, Falkovich \& Lebedev 1997, Chertkov, Kolokolov \& Vergassola
1997, Balkovsky et al.~1997, Balkovsky \& Falkovich 1998). For
incompressible turbulence the theory predicts velocity pdf's which are
mainly Gaussian with only minor enhancement at the far ends of the
tails. The distribution of velocity {\em differences} (between
locations in the system separated by a given shift vector $\Delta
\vec{r}$) is expected to deviate considerably from being normal and is
likely to resemble an exponential. This finding is supported by a
variety of experimental and numerical determinations (e.g.~Kida \&
Murakami 1989, Vincent \& Meneguzzi 1991, Jayesh \& Warhaft 1991, She
1991, She, Jackson \& Orszag 1991, Cao, Chen, \& She 1996, Vainshtein
1997, Lamballais, Lesieur, \& M{\'e}tais 1997, Machiels \& Deville
1998). Compressible turbulence has remained to be too complex for a
satisfying mathematical analysis.

\subsection{PDF's of Observable Quantities}
\label{subsec:pdf}
It is not clear how to relate the analytical work on incompressible
turbulence to molecular clouds. In addition to the fact that
interstellar turbulence is highly supersonic and self-gravitating,
there are also observational limitations. Unlike the analytical
approach or numerical simulations, molecular cloud observations allow
access only to dimensionally reduced information.  Velocity
measurements are possible only along the line-of-sight, and the
spatial structure of a cloud is only seen in projection onto the plane
of the sky, i.e.~as variations of the column density. Although some
methods can yield information about the 3-dimensional spatial
structure of the cloud (see Stutzki \& G{\"u}sten 1990, Williams,
De~Geus, \& Blitz 1994), the result is always model dependent and
equivocal (see also Ballesteros-Paredes, V{\'a}zquez-Semadeni, \&
Scalo 1999).

A common way of obtaining knowledge about the velocity structure of
molecular clouds is to study individual line profiles at a large
number of various positions across the cloud. In the optical thin case
line shapes are in fact histograms of the radial velocities of gas
sampled along the telescope beam.  Falgarone \& Phillips (1990) and
Falgarone et al.\ (1994) showed that line profiles constructed from
high-sensitivity CO maps exhibit non-Gaussian wings and attributed
this to turbulent intermittency (see also Falgarone et al.\ 1998 on
results from the IRAM-key project).  Dubinski, Narayan, \& Phillips
(1995) demonstrated that non-Gaussian line profiles can be produced
from {\em any} Gaussian random velocity field if variance effects
become important (which is always the case for very steep or truncated
power spectra). They concluded that non-Gaussian line profiles do not
provide clear evidence for intermittency.

Another method of inferring properties of the velocity distribution in
molecular clouds is to analyze the pdf of line centroid velocities
obtained from a large number of individual measurements scanning the
entire projected surface area of a cloud (Miesch \& Scalo 1995, Lis et
al.~1998, Miesch et al.~1998). Each line profile (i.e. the pdf {\em
along} the line-of-sight) is collapsed into one single number, the
centroid velocity, and then sampled {\em perpendicular} to the
line-of-sight. Hence, the two functions differ in the direction of the
sampling and in the quantity that is considered. A related statistical
measure is the pdf of centroid velocity increments, it samples the
velocity differences between the centroids for line measurements which
are offset by a given separation.  The observational advantage of
using centroid and increment pdf's is, that the line measurements can
typically be taken with lower sensitivity as only the centroid has to
be determined instead of the detailed line shape. These measures are
also less dependent on large-scale systematic motions of the cloud and
they are less effected by line broadening due to the possible presence
of warm dilute gas. However, to allow for a meaningful analysis of the
pdf's especially in the tails, the number of measurements needs to be
very large and should not be less than about 1000. In order to sample
the entire volume of interstellar clouds, the molecular lines used to
obtain the pdf's are optically thin. I follow this approach in the
present investigation and use a mass-weighted velocity sampling along
the line-of-sight to determine the line centroid. This zero-opacity
approximation does not require any explicit treatment of the radiation
transfer process.

The observed pdf's are obtained from {\em averaged} quantities (from
column densities or line centroids). To relate these observational
measures to quantities relevant for turbulence theory, i.e.~to the
full 3-dimensional pdf, numerical simulations are necessary as only
they allow unlimited access to all variables in phase space. A first
attempt to do this was presented by Lis et al.~(1996, 1998) who
analyzed a simulation of mildly supersonic decaying hydrodynamic
turbulence by Porter et al.~(1994). Since their model did neither
include self-gravity nor consider flows at high Mach number or
mechanisms to replenish turbulence, the applicability to the
interstellar medium remained limited.  This fact prompts the current
investigation which extends the previous ones by calculating {\em
highly supersonic flows}, and by including {\em self-gravity} and a
{\em turbulent driving scheme}.  The current study does not consider
magnetic fields. Their influence on the pdf's needs to be addressed
separately.  However, the overall importance of magnetic fields and
MHD waves on the dynamical structure of molecular clouds may not be
large. The energy associated with the observed fields is of the order
of the (turbulent) kinetic energy content of molecular clouds
(Crutcher 1999).  Magnetic fields cannot prevent the decay of
turbulence (e.g.\ Mac Low et al.~1998) which implies the presence of
external driving mechanisms. These energy sources replenish the
turbulent cascade and may excite MHD waves explaining the inferred
equipartition between turbulent and magnetic energies.

\subsection{Statistical Definitions}
\label{subsec:definitions} 
 The one-point probability distribution function $f(x)$ of a variable
$x$ is defined such that $f(x)dx$ measures the probability for the
variable to be found in the interval $[x,x+dx]$. The {\em density pdf}
($\rho$-pdf) discussed in this paper is obtained from the local
density associated with each SPH particle. It is basically the
normalized histogram summed over all particles in the simulation,
i.e.\ a mass-weighted sampling procedure is applied.  The {\em
line-of-sight velocity centroid pdf} ($v$-pdf) is more complicated to
compute.  The face of the simulated cube is divided into $64^2$
equal-sized cells. For each cell, the line profile is computed by
sampling the normal (line-of-sight) velocity component of all gas
particles that are projected into that cell. The line centroid is
determined as the abscissa value of the peak of the distribution.
This procedure corresponds to the formation of optically thin lines in
molecular clouds, where all molecules within a certain column through
the clouds contribute equally to the shape and intensity of the line.
To reduce the sampling uncertainties, this procedure is repeated with
the location of the cells shifted by half a cell size in each
direction. Altogether about 20$\,$000 lines contribute to the
pdf. This is procedure is repeated for line-of-sights along all three
system axes to identify projection effects. The {\em line centroid
increment pdf} ($\Delta v$-pdf) is obtained in a similar
fashion. However, the sampled quantity is now the velocity {\em
difference} between line centroids obtained at two distinct locations
separated across the face of the cloud by a fixed shift vector $\Delta
\vec{r}$.  The $\Delta v$-pdf for a spatial lag $\Delta r$ is obtained
as azimuthal average, i.e.~as superposition of all individual pdf's
with shift vectors of length $\Delta r$.  

Also statistical moments of the distribution can be used to quantify
the spread and shape of pdf's. For the current analysis I use the
first four moments.  Mean value $\mu$ and standard deviation $\sigma$
(the 1.\ and 2.\ moments) quantify the location and the width of the
pdf and are given in units of the measured quantity. The third and
fourth moments, skewness $\theta$ and kurtosis $\kappa$, are
dimensionless quantities characterizing the shape of the
distribution. The skewness $\theta$ describes the degree of asymmetry
of a distribution around its mean. The kurtosis $\kappa$ measures the
relative peakedness or flatness of the distribution. I use a
definition where $\kappa=3$ corresponds to a normal
distribution. Smaller values indicate existence of a flat peak
compared to a Gaussian, larger values point towards a stronger peak or
equivalently towards the existence of prominent tails in the
distribution. A pure exponential results in $\kappa=6$.  Gaussian
random fields are statistically fully determined by their mean value
and the 2-point correlation function, i.e.~by their first two moments,
$\mu$ and $\sigma$. All higher moments can be derived from those. The
2-point correlation function is equivalent to the power spectrum in
Fourier space (e.g.\ Bronstein \& Semendjajew 1979).

Besides using moments there are other possibilities of characterizing
a distribution. Van den Marel \& Franx (1993) and Dubinski et
al.~(1995) applied Gauss-Hermite expansion series to quantify
non-normal contributions in line profiles.  A more general approach
has been suggested by Vio et al.~(1994), who discuss alternatives to
the histogram representation of pdf's. However, as astrophysical data
sets typically {\em are} histograms of various types and as histograms
are the most commonly used method to describe pdf's, this approach is
also adopted here.

\section{The Numerical Model}
\label{sec:model}

\subsection{SPH in Combination with GRAPE}
SPH ({\em smoothed particle hydrodynamics}) is a particle-based scheme
to solve the equations of hydrodynamics.  The fluid is represented by
an ensemble of particles, each carrying mass, momentum, and
hydrodynamic properties.  The time evolution of the fluid is
represented by the time evolution of the particles, governed by the
equations of motion which are supplemented by a prescription to modify
the hydrodynamic properties. At any location these properties are
obtained by averaging over an appropriate set of neighboring
particles.  Excellent overviews over the method provide the reviews by
Benz (1990) and Monaghan (1992).  For the current study I use SPH
because it is intrinsically Lagrangian and because it is able to
resolve very high density contrasts.  Another reason for choosing SPH
is the possibility to use it in combination with the special-purpose
hardware device GRAPE (Sugimoto et al.~1990, Ebisuzaki et al.~1993;
and also Umemura et al.~1993, Steinmetz 1996). This allows
calculations at supercomputer level on a normal workstation.

The code is based on a version originally developed by Benz (1990),
and is used with a standard description of a von~Neumann-type
artificial viscosity (Monaghan \& Gingold 1983) with the parameters
$\alpha_v = 1$ and $\beta_v = 2$ for the linear and quadratic
terms. The system is subject to periodic boundary conditions (Klessen
1997) and is integrated in time using a second-order
Runge-Kutta-Fehlberg scheme, allowing individual time steps for each
particle. Furthermore, the smoothing volume over which hydrodynamic
quantities are averaged in the code is freely adjustable in space and
time such that the number of neighbors for each particle remains
approximately fifty.  When including self-gravity, regions with masses
exceeding the Jeans limit become unstable and collapse. Once a
highly-condensed core has formed in the center of a collapsing gas
clump, that core is substituted by a `sink' particle (Bate, Bonnell,
\& Price 1995) which inherits the combined masses, linear and `spin'
angular momenta of the particles it replaces. It also has the ability
to accrete further SPH particles from its infalling gaseous envelope.

For simulations of turbulent flows one also has to take into account
that an explicit viscosity term is introduced in the SPH method. This
fact demands attention when studying dissipative processes, especially
in the subsonic regime. The current study focuses on the properties of
highly supersonic turbulent flows. In this regime, direct comparison
between SPH and grid-based methods has proven the close correspondence
of both methods (Mac~Low et al.~1998, Klessen, Heitsch \& Mac~Low
2000).  If one bears the above caveats in mind, the SPH method
calculates the time evolution of gaseous systems very reliably and
accurately, and offers large spatial and dynamical flexibility.

\subsection{Models}
\label{subsec:models}
The numerical models discussed here describe isothermal gas. The
hydrodynamic equations are extended to include self-gravity (in Sec.'s
\ref{sec:decay-with-gravity} and \ref{sec:driven-with-gravity}) and to
incorporate a random turbulent driving mechanism (in Sec.\
\ref{sec:driven-with-gravity}).  All physical constants are set to
unity. The same applies to mass and length scales, i.e.\ the total
mass is $M=1$ and the simulated volume is the cube $[-1,+1]^3$. The
mean density is thus $\rho = 1/8$. The initial configuration of all
dynamical systems discussed in this paper is a homogeneous gas
distribution with a Gaussian velocity field.  Without turbulence, the
time evolution depends on one parameter, the {\em ratio} between
internal and gravitational energy, $\alpha \equiv \epsilon_{\rm
int}/|\epsilon_{\rm pot}|$. This quantity can be interpreted as
dimensionless temperature and determines the number of thermal Jeans
masses contained in the system. Molecular clouds are characterized by
line widths which largely exceed the thermal broadening.  The
evolution away from the homogeneous initial state is thus strongly
influenced by the adopted initial velocity distribution and depends on
whether turbulence is decaying or driven. Large turbulent kinetic
energy can considerable slow down or even prevent the collapse of
thermally Jeans unstable gas. The situation is very complex and
depends on the shape and strength of the turbulent velocity spectrum
(Klessen et al.~2000; see also see Bonazzola et al.~1992 and
V{\'a}zquez-Semadeni \& Gazol 1995 for an analytical approach).

To generate and maintain turbulent flows Gaussian velocity fields are
introduced. The spatial variations of each component of the velocity
vector $\vec{v}$ are described as superpositions of plane waves with
wave numbers $\vec{k} = (k_x, k_y, k_z)$, where the phase of each wave
is random and sampled from a uniform distribution in the interval
$[0,2\pi[$. Also the amplitude is random, but selected from a Gaussian
distribution centered on zero and with a width determined by the power
spectrum $P(k) = A_k k^{\alpha}$. Gaussian fields are isotropic and
only depend on the absolute value of the wave vector $k=|\vec{k}|$.
Only waves in the range $1 \leq k \leq k_{\rm max}$ are
considered. For large cut-off wave numbers $k_{\rm max}$ the Gaussian
statistics is very well sampled. If only very few modes are used to
generate the field, variance effects become strong and individual
realizations of the field can deviate significant from the ensemble
average (see Sec.~\ref{sec:init}).  The field is then transformed back
into real space and the resulting velocities are assigned to
individual SPH particles using the `cloud-in-cell' scheme (Hockney \&
Eastwood 1988). For the initial field, all velocities are multiplied
by the appropriate factor to reach the desired rms Mach number of the
flow. In case of driven turbulence, this velocity field is also used
to `kick' the SPH particles at every time step such that a constant
level of kinetic energy is maintained (see Mac~Low 1999)

\section{PDF's from Gaussian Velocity Fluctuations}
\label{sec:init}

Variance effects in poorly sampled Gaussian velocity fields can lead
to considerable {\em non}-normal contributions to the $v$- and $\Delta
v$-pdf's.  If a random process is the result of sequence of
independent events (or variables), then in the limit of large numbers,
its distribution function will be a Gaussian around some mean value.
However, only the properties of a large {\em ensemble} of Gaussian
fields are determined in a statistical sense. Individual realizations
may exhibit considerable deviations from the mean. The effect is
strongest when only few (spatial) modes contribute to the field or,
almost equivalently, when the power spectrum falls off very
steeply. In this case, most kinetic energy is in large-scale motions.

\begin{figure}[h]
\unitlength1.0cm
\begin{picture}(18,12.5)
\put( -2.25,-12.75){\epsfbox{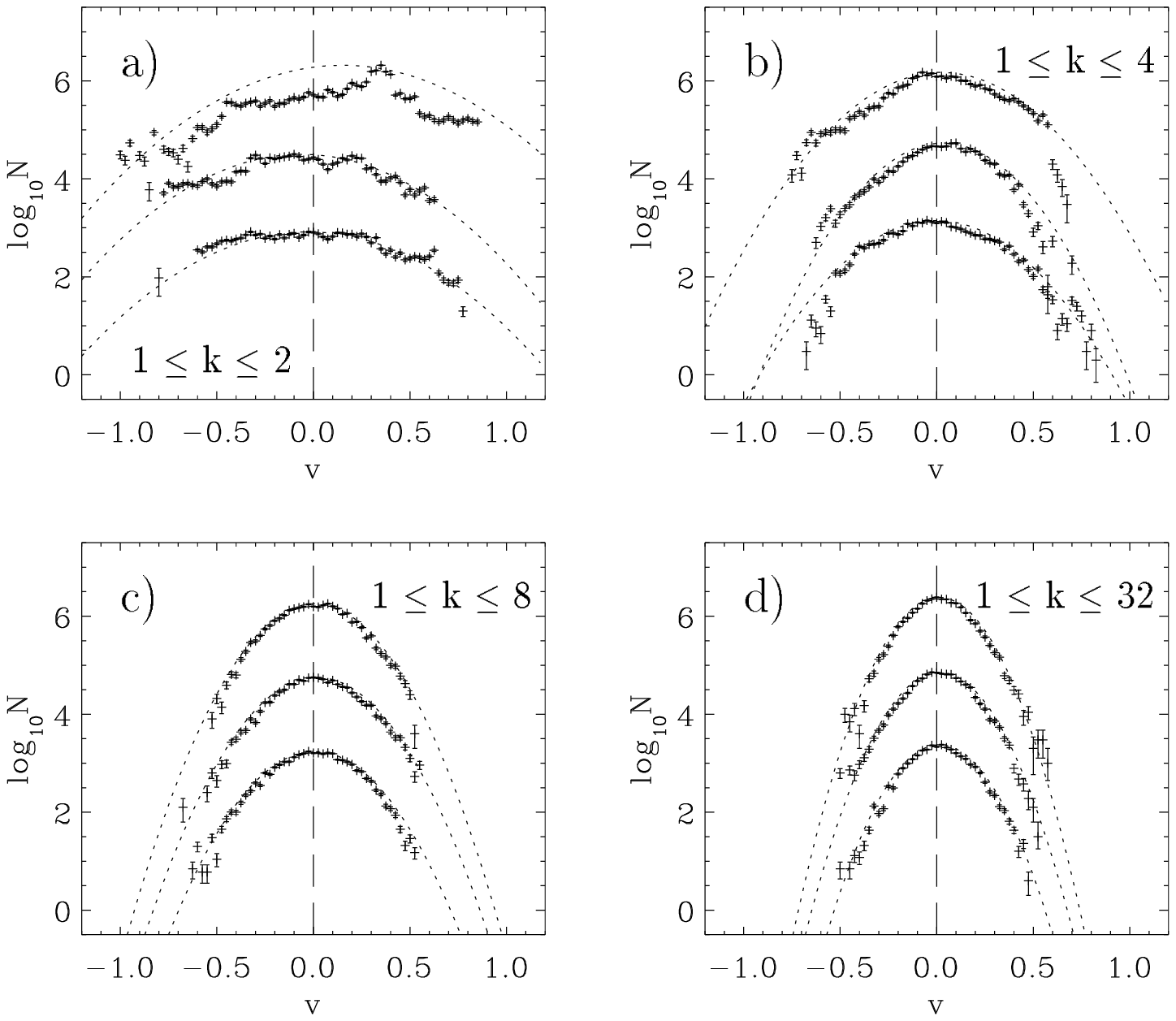}}
\end{picture}
\caption{\label{fig:v-pdf-init} Pdf's of line centroids for a
homogeneous gaseous medium with Gaussian velocity field. The power
spectrum is $P(k) = {\rm const.}$ with wave numbers in the intervals
(a) $1\leq k\leq2$, (b) $1\leq k\leq 4$, (c) $1\leq k\leq 8$, to (d)
$1\leq k\leq 32$. All other modes are suppressed. Each figure plots
pdf's of the $x$-, $y$-, and $z$-component of the velocity offset by
$\Delta \log_{10} N = 1.5$ (lowest, middle, and upper distribution,
respectively). The length of the error bars
is determined by the square root of the numbers of entries per
velocity bin. The Gaussian fit from the first two moments is shown
with dotted lines.}
\end{figure}

 This is visualized in Fig.~\ref{fig:v-pdf-init}, it shows $v$-pdf's
for homogeneous gas (sampled by $64^3$ SPH particles placed on a
regular grid) with Gaussian velocity fields with power spectra $P(k) =
{\rm const.}$ which are truncated at different wave numbers $k_{\rm
max}$ ranging from (a) $k_{\rm max}=2$ to (d) $k_{\rm max}=32$. Each
realization is scaled such that the rms velocity dispersion is
$\sigma_v = 0.5$.  The figure displays the pdf's for the $x$-, $y$-,
and $z$-component of the velocity. The pdf's of the strongly truncated
spectrum (Fig.~\ref{fig:v-pdf-init}a) do not at all resemble normal
distributions. The Gaussian statistics of the field is very badly
sampled with only very few modes.  Note that the pdf's of the same
field may vary considerably for different velocity {\em components},
i.e.~for different {\em projections}.  With the inclusion of larger
number of Fourier modes this situation improves, and in
Fig.~\ref{fig:v-pdf-init}d the pdf's of all projections sample the
expected Gaussian distribution very well.

\begin{figure}[h]
\unitlength1.0cm
\begin{picture}(18,12.5)
\put( -2.25,-12.75){\epsfbox{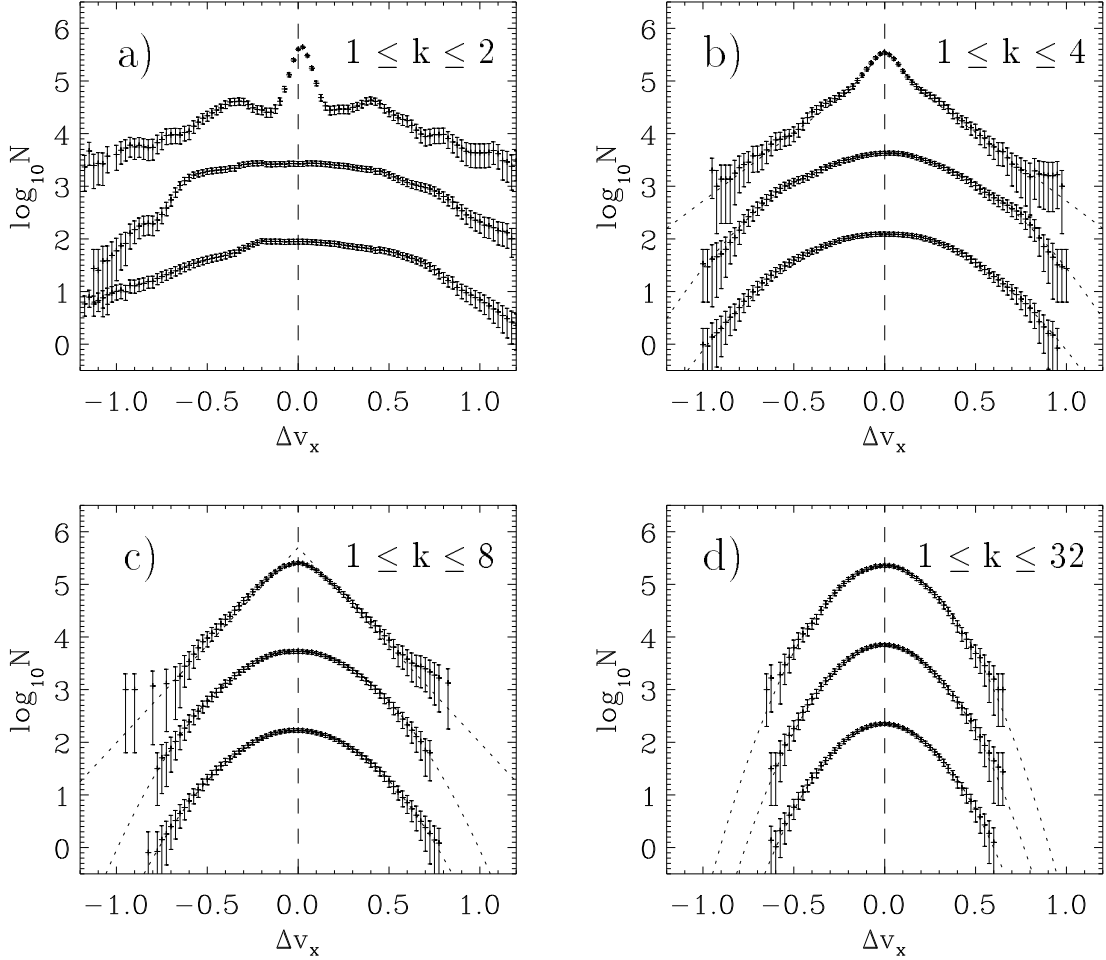}}
\end{picture}
\caption{\label{fig:dv-pdf-init} Pdf's of line centroid increments for
the same systems as in Fig.~\ref{fig:v-pdf-init}: (a) $1\leq k\leq2$,
(b) $1\leq k\leq 4$, (c) $1\leq k\leq 8$, to (d) $1\leq k\leq
32$. Each plot shows the distribution of centroid velocity differences
between locations separated by the distance $\Delta r$ --- upper
curve: $\Delta r = 1/32$, middle curve: $\Delta r = 10/32$, and lower
curve: $\Delta r = 30/32$. Only the velocity component for the
line-of-sight parallel to the $x$-axis is considered. Again, the
dotted lines represent the best fit Gaussian, except for the upper
curve in (b) and (c) where the best exponential fit is shown.  }
\end{figure}

A similar conclusion can be derived for $\Delta v$-pdf. This measure
is even more sensitive to deviations from Gaussian statistics.
Figure~\ref{fig:dv-pdf-init} plots the $\Delta v$-pdf's for the same
sequence of velocity fields. For brevity, only the line-of-sight
component parallel to the $x$-axis is considered. Furthermore, from
the sequence of possible $\Delta v$-pdf's (defined by the spatial lag
$\Delta r$) only three are shown, at small ($\Delta r=1/32$, upper
curve), medium ($\Delta r=10/32$, middle curve), and large spatial
lags ($\Delta r=30/32$, upper curve). Sampling the Gaussian field with
only two modes (Fig.~\ref{fig:dv-pdf-init}a) is again insufficient to
yield increment pdf's of normal shape. The velocity field is very
smooth, and the line centroid velocity difference between neighboring
cells is very small. Hence, for $\Delta r=1/32$ the pdf is dominated
by a distinct central peak at $\Delta v = 0$.  The tails of the
distribution are quite irregularly shaped. The situation becomes
`better' when sampling increasing distances, as regions of the fluid
separated by larger $\Delta r$ are less strongly correlated in
velocity. For $\Delta r=10/32$ and $\Delta r=30/32$ the pdf's follow
the Gaussian distribution more closely although irregularities in the
shapes are still present. In Fig.'s \ref{fig:dv-pdf-init}b and c the
$\Delta v$-pdf's for medium to large lags are very well fit by
Gaussians. Deviations occur only at small $\Delta r$, the pdf's are
exponential (and the distribution for $k_{\rm max} = 4$ is still a bit
cuspy). Finally, Fig.~\ref{fig:dv-pdf-init}d shows the three $\Delta
v$-pdf's for the case where all available spatial modes contribute to
the velocity field ($1\leq k \leq 32$). The pdf's follow a Gaussian
for all spatial lags.

\begin{figure}[h]
\unitlength1.0cm
\begin{picture}(18,6.55)
\put( -2.25,-18.20){\epsfbox{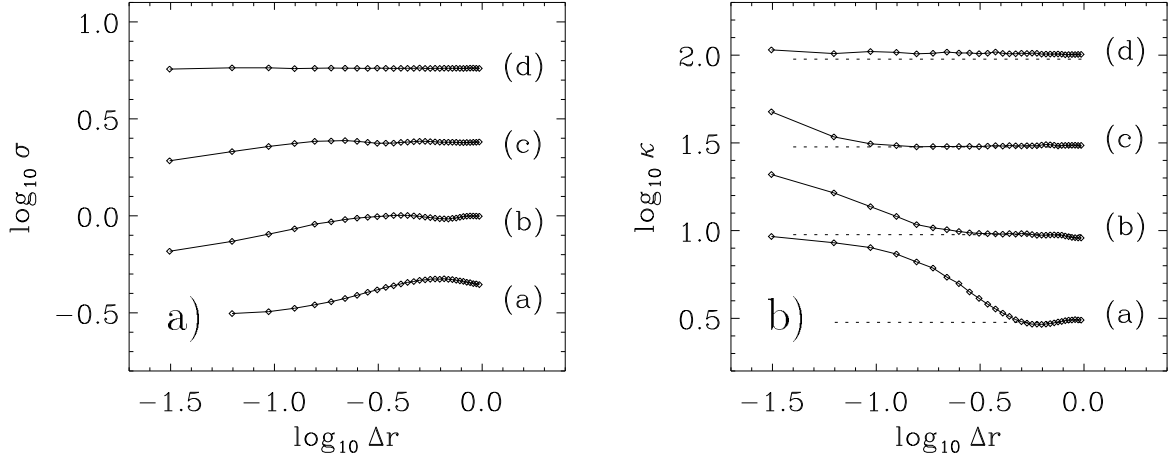}}
\end{picture}
\caption{\label{fig:mom-init} (a) The second, dispersion $\sigma$, and
(b) the fourth moment, kurtosis $\kappa$, of the distribution of
velocity increments displayed in Fig.~\ref{fig:dv-pdf-init} as
function of spatial lag $\Delta r$. The letters on the right-hand
sight indicate correspondence to the previous figure. Each plot is
offset by $\Delta \log_{10} \sigma=0.5$ and $\Delta \log_{10} \sigma =
0.5$, and in (b) the horizontal dotted line indicates the value for a
Gaussian $\kappa = 3$ ($\log_{10} \kappa = 0.48$). }
\end{figure}

This behavior is also seen in the variation of the moments of the
distribution as function of the spatial lag $\Delta r$.  Applied to
the above sequence of Gaussian velocity fields,
Fig.~\ref{fig:mom-init} displays the dispersion $\sigma$ and the
kurtosis $\kappa$ of the distribution. The corresponding models are
indicated at the right hand side of each plot. The width of the
distribution, as indicated by the dispersion $\sigma$
(Fig.~\ref{fig:mom-init}a), typically grows with increasing $\Delta
r$, reflecting the relative peakedness of the distribution at small
lags. For example, the distribution (a) yields a slope of ~0.3 in the
range $-0.6\leq \log_{10} \Delta r \leq -0.4$, and (b) leads to a
value of 0.2 in relatively large interval $-1.5 \leq \log_{10} \Delta
r \leq -0.5$. The effect disappears for the better sampled
fields. Typical values for that slope in observed molecular clouds are
$-0.3$ to $-0.5$ (Miesch et al.~1998).\footnote{Note, that Miesch et
al.~(1998) are plotting the function $\sigma^2$ versus the spatial lag
$\Delta r$. For a comparison with the present study, their numbers
have to be divided by a factor of two. Furthermore, they use a
relatively narrow range of $\Delta r$-values to compute the slope of
the function; larger intervals would on average tend to decrease these
values (see their Fig.~14). In addition, Miesch et al.~(1998) applied
spatial filtering to remove large-scale velocity gradients in the
clouds. These would lead to steeper slopes.  The fact that in the
present study the functions $\sigma$ and $\kappa$ level out for large
spatial lags $\Delta r$ is a consequence of the periodic boundary
conditions which do not allow for large-scale
gradients.\label{comment}} A direct measure of the peakedness of the
distribution is its fourth moment, the kurtosis $\kappa$
(Fig.~\ref{fig:mom-init}b).  At small lags $\Delta r$, clearly the
pdf's of model (a) are more strongly peaked than exponential ($\kappa
=6$). Comparing the entire sequence reveals again the tendency of the
pdf's to become Gaussian at decreasing $\Delta r$ with increasing
number of modes considered in the construction of the velocity field.

Taking all together, it is advisable to consider conclusions about
interstellar turbulence derived from solely analyzing one-point
probability distribution functions from molecular clouds with
caution. Similar to what has been shown by Dubinski et al.~(1995) for
molecular line profiles, deviations from the regular Gaussian shape
found in $v$- and $\Delta v$-pdf's need not be the signpost of
turbulent intermittency.  Gaussian velocity fields which are dominated
by only a small number of modes (either because the power spectrum
falls off steeply towards larger wave numbers, or because small wave
length distortions are cut away completely) will lead to very similar
distortions. In addition, the properties of the pdf may vary
considerably between different projections. The same velocity field
may lead to smooth and Gaussian pdf's for one velocity component,
whereas another projection may result in strong non-Gaussian wings
(see also Fig.~\ref{fig:v-pdf-xyz-decay-with-gravity}).

\section{Analysis of Decaying Supersonic Turbulence without Self-Gravity}
\label{sec:decay-non-grav}

In this section the pdf's of freely decaying initially highly
supersonic turbulence without self-gravity are discussed.  They are
calculated from an SPH simulation with $350\,000$ particles (Mac~Low
et al.~1998, model G). Initially the system is homogeneous with a
Gaussian velocity distribution with $P(k) = {\rm const.}$ in the
interval $1 \leq k \leq 8$. The rms Mach number of the flow is $M=5$.

After the onset of the hydrodynamic evolution the flow quickly becomes
fully turbulent resulting in rapid dissipation of kinetic energy. The
energy decay is found to follow a power law $t^{-\eta}$ with exponent
$\eta = 1.1 \pm 0.004$.  The overall evolution can be subdivided into
several phases. The first phase is very short and is defined by the
transition of the initially Gaussian velocity field into fully
developed supersonic turbulence. It is determined by the formation of
the first shocks which begin to interact with each other and build up
a complex network of intersecting shock fronts. Energy gets transfered
from large to small scales and the turbulent cascade builds up. The
second phase is given by the subsequent self-similar evolution of the
network of shocks. Even though individual features are transient, the
overall properties of this network change only slowly. In this phase
of highly supersonic turbulence the loss of kinetic energy is
dominated by dissipation in shocked regions. In the transsonic regime,
i.e.~the transition from highly supersonic to fully subsonic flow,
energy dissipation in vortices generated by shock interactions becomes
more and more important. Only the strongest shocks remain in this
phase. Surprisingly, the energy decay law does not change during this
transition. It continues to follow a power law with exponent $\eta
\approx 1$.  In the subsonic phase the flow closely resembles
incompressible turbulence.  Its properties are similar to those
reported from numerous experiments and simulations (e.g.~Porter et
al.~1994, Lesieur 1997, Boratav et al.~1997). The simulation is
stopped at $t=20.0$ when the flow has decayed to a rms Mach number of
$M=0.3$.  Since the energy loss rate follows a power law, the duration
of each successive phase grows.

\begin{figure}[h]
\unitlength1.0cm
\begin{picture}(18,7.3)
\put( -2.25,-18.20){\epsfbox{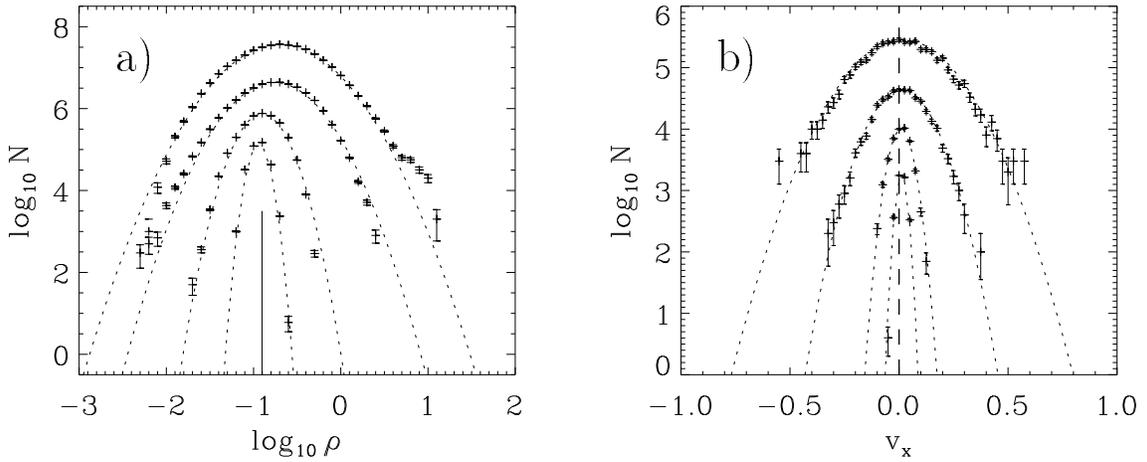}}
\end{picture}
\caption{\label{fig:rho-v-pdf-decay-no-gravity}
Pdf's of (a) density and of (b) centroid velocities for the
line-of-sight being parallel to the $x$-axis of the system. The pdf's
are obtained at four different phases of the dynamical evolution of
the system (see the main text), at $t=0.2$ (upper curves), at $t=0.6$
(second curve from the top), at $t=3.5$ (third curve), and at $t=20.0$
(lowest curve). These times correspond to Mach numbers  $M=5.0$,
$M=2.5$, $M=1.0$, and $M=0.3$, respectively. For each distribution,
the best-fit Gaussian is indicated using dotted lines.
}
\end{figure}

This sequence of evolutionary stages is seen in the pdf's of the
system. One noticeable effect is the decreasing width of the
distribution functions as time progresses. As the kinetic energy
decays the available range of velocities shrinks. This not only leads
to `smaller' $v$- and $\Delta v$-pdf's, but also to a smaller
$\rho$-pdf since compressible motions lose influence and the system
becomes more homogeneous. This is indicated in
Fig.~\ref{fig:rho-v-pdf-decay-no-gravity}, it displays (a) the
$rho$-pdf and (b) $v$-pdf at the following stages of the dynamical
evolution (from top to bottom): Shortly after the start, at $t=0.2$
when the first shocks occur, then at $t=0.6$ when the network of
interacting shocks is established and supersonic turbulence is fully
developed, during the transsonic transition at $t=3.5$, and finally at
$t=20.0$ when the flow has progressed into the subsonic regime. The
rms Mach numbers at these stages are $M=5.0$, $M=2.5$, $M=1.0$, and
$M=0.3$, respectively. The density pdf always closely follows a
log-normal distribution, i.e.~it is Gaussian in the {\em logarithm} of
the density. Also the distribution of line centroids at the four
different evolutionary stages of the system is best described by a
Gaussian with only minor deviations at the far ends of the velocity
spectrum.

\begin{figure}[h]
\unitlength1.0cm
\begin{picture}(18,12.5)
\put( -2.25,-12.75){\epsfbox{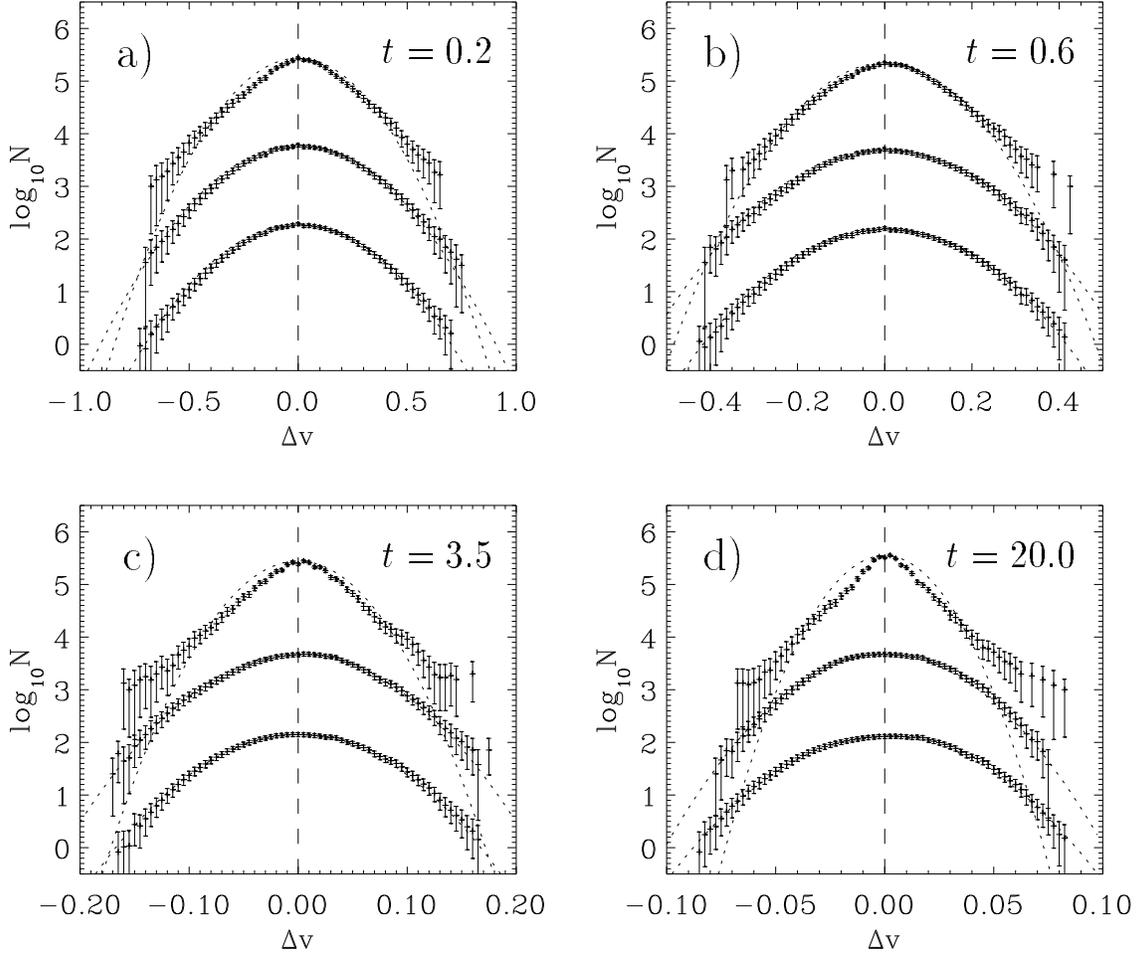}}
\end{picture}
\caption{\label{fig:dv-pdf-decay-no-gravity} Pdf's of the
$x$-component of the centroid velocity increments for three spatial
lags: upper curve -- $\Delta r = 1/32$, middle curve -- $\Delta r =
10/32$, and lower curve -- $\Delta r = 30/32$. As in
Fig.~\ref{fig:rho-v-pdf-decay-no-gravity}, the pdf's are obtained at
(a) $t=0.2$ , (b) $t=0.6$, (c) $t=3.5$, and (d) $t=20.0$. The Gaussian
fits are again indicated by dotted lines.  }
\end{figure}

For the same points in time, Fig.~\ref{fig:dv-pdf-decay-no-gravity}
shows the $\Delta v$-pdf's for $x$-component of the velocity. The
displayed spatial lags are selected in analogy to
Fig.~\ref{fig:dv-pdf-init}.  Note the different velocity scaling in
each plot reflecting the decay of turbulent energy as the system
evolves in time.  Throughout the entire sequence, spatial lags larger
than about 10\% of the system size always lead to $\Delta v$-pdf's
very close to Gaussian shape (the middle and lower curves).
Considerable deviations occur only at small spatial lags (the upper
curves). For those, the increment pdf's exhibit exponential wings
during all stages of the evolution. When scaling the pdf's to the same
width, the distribution in the subsonic regime (d) appears to be more
strongly peaked than during the supersonic or transsonic phase (a --
c). There, the central parts of the pdf's are still reasonably well
described by the Gaussian obtained from the first two moments, whereas
in (d) the peak is considerably narrower, or vice versa, the tails of
the distribution are more pronounced.

These results can be compared with the findings by Lis et
al.~(1998). They report increment pdf's for three snapshots of a
high-resolution hydrodynamic simulation of decaying mildly super-sonic
turbulence performed by Porter et al.~(1994). They analyze the system
at three different times corresponding to rms Mach numbers of
$M\approx 0.96$, $M\approx 0.88$, and $M\approx 0.52$. Their first two
data sets thus trace the transition from supersonic to subsonic flow
and are comparable to phase (c) of the current model; their last data
set corresponds to to phase (d).  In the transsonic regime both
studies agree: Lis et al.~(1998) report enhanced tails in the
increment pdf's for the smallest spatial lags which they considered
and near Gaussian distributions for larger lags (however, the largest
separation they study is about 6\% of the linear extent of the
system). In the subsonic regime, Lis et al.~(1998) find near Gaussian
pdf's for very small spatial lags ($<1$\%), but extended wings in the
pdf's for lags of 3\% and 6\% of the system size. They associate this
with the `disappearance' of large-scale structure. Indeed, their
Fig.~7 exhibits a high degree of fluctuations on small scales which
they argue become averaged away when considering small spatial lags in
the $\Delta v$-pdf. Comparing the pdf with spatial lags of 3\% (upper
curves in Fig.~\ref{fig:dv-pdf-decay-no-gravity}, compared to the
pdf's labeled with $\Delta = 15$ in Lis et al.~1998) both studies come
to the same result. At these scales the $\Delta v$-pdf's tend to
exhibit more pronounced wings in the subsonic regime as in the
supersonic regime. The SPH calculations reported here do not allow for
a meaningful construction of $\delta v$-pdf's for $\Delta r <
3$\%. The Gaussian behavior of pdf's for very small spatial lags
reported by Lis et al.~(1998) therefore cannot be examined.  However,
neither of the purely hydrodynamic simulations lead to pdf's that
are in good agreement with the observations. Observed pdf's typically
are much more centrally peaked at small spatial separation (see e.g.\
Fig.~4 in Lis et al~1998 and Miesch et al.~1998).

\begin{figure}[h]
\unitlength1.0cm
\begin{picture}(18,19.0)
\put( -2.25,-9.0){\epsfbox{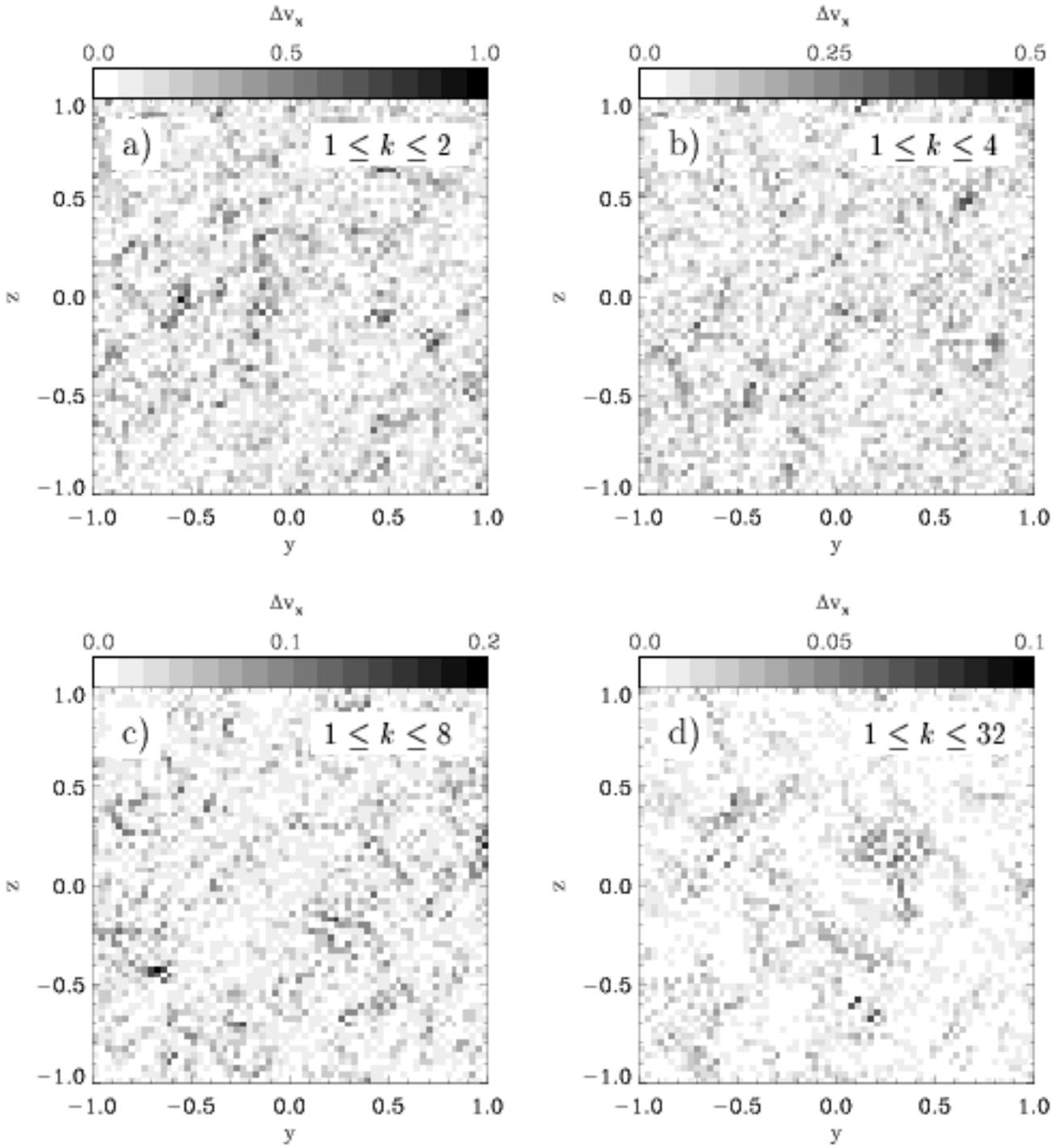}}
\end{picture}
\caption{\label{fig:dv-array-decay-no-gravity}
2-dimensional distribution (in the $yz$-plane) of centroid
increments for velocity profiles along the $x$-axis of the system
between locations separated by a vector lag $\Delta \vec{r} =
(1/32,1/32)$. Analog to the previous figures, the data are displayed
for times (a)  $t=0.2$, (b) $t=0.6$, (c) $t=3.5$, and (d) $t=20.0$.
The magnitude of the velocity increment $\Delta v_{x}$ is indicated at
the top of each plot;  note the different scaling. 
}
\end{figure}

Figure \ref{fig:dv-array-decay-no-gravity} shows the spatial
distribution of centroid velocity differences between cells separated
by a vector lag of $\Delta \vec{r} = (1/32,1/32)$ (i.e.~between
neighboring cells along the diagonal). Data are obtained at the same
times as above.  Each figure displays the array of the absolute values
of the velocity increments $\Delta v_x$ in linear scaling as indicated
at the top. Note the decreasing velocity range reflecting the decay of
turbulent energy.  The distribution of $\Delta v_x$ appears random,
there is no clear indication for coherent structures.  This is
corresponds to most observations. Miesch et al.~(1998) find for their
sample of molecular clouds that high-amplitude velocity differences
for very small spatial lags typically are well distributed resulting
in a `spotty' appearance. Note, however, that using azimuthal
averaging Lis et al.~1998 report the finding of filamentary structures
for the $\rho$-Ophiuchus cloud. Altogether, filamentary structure is
difficult to define and a mathematical thorough analysis is seldomly
performed (for an astrophysical approach see Adams \& Wiseman 1994,
for a discussion of the filamentary vortex structure in incompressible
turbulence consult Frisch 1995 or Lesieur 1997). The visual inspection
of maps is often misleading and influenced by the parameters used to
display the image.  Larger velocity bins for instance tend to produce
a more `filamentary' structure than very fine sampling of the velocity
structure. Further uncertainty may be introduced by the fact that
molecular clouds are only seen in one projection as the signatures of
the dynamical state of the system can strongly depend on the viewing
angle.

\section{Analysis of Decaying Turbulence with Self-Gravity}
\label{sec:decay-with-gravity}
\begin{figure}[h]
\unitlength1.0cm
\begin{picture}(18,19.1)
\put( -0.90,-4.5){\epsfysize=27cm\epsfbox{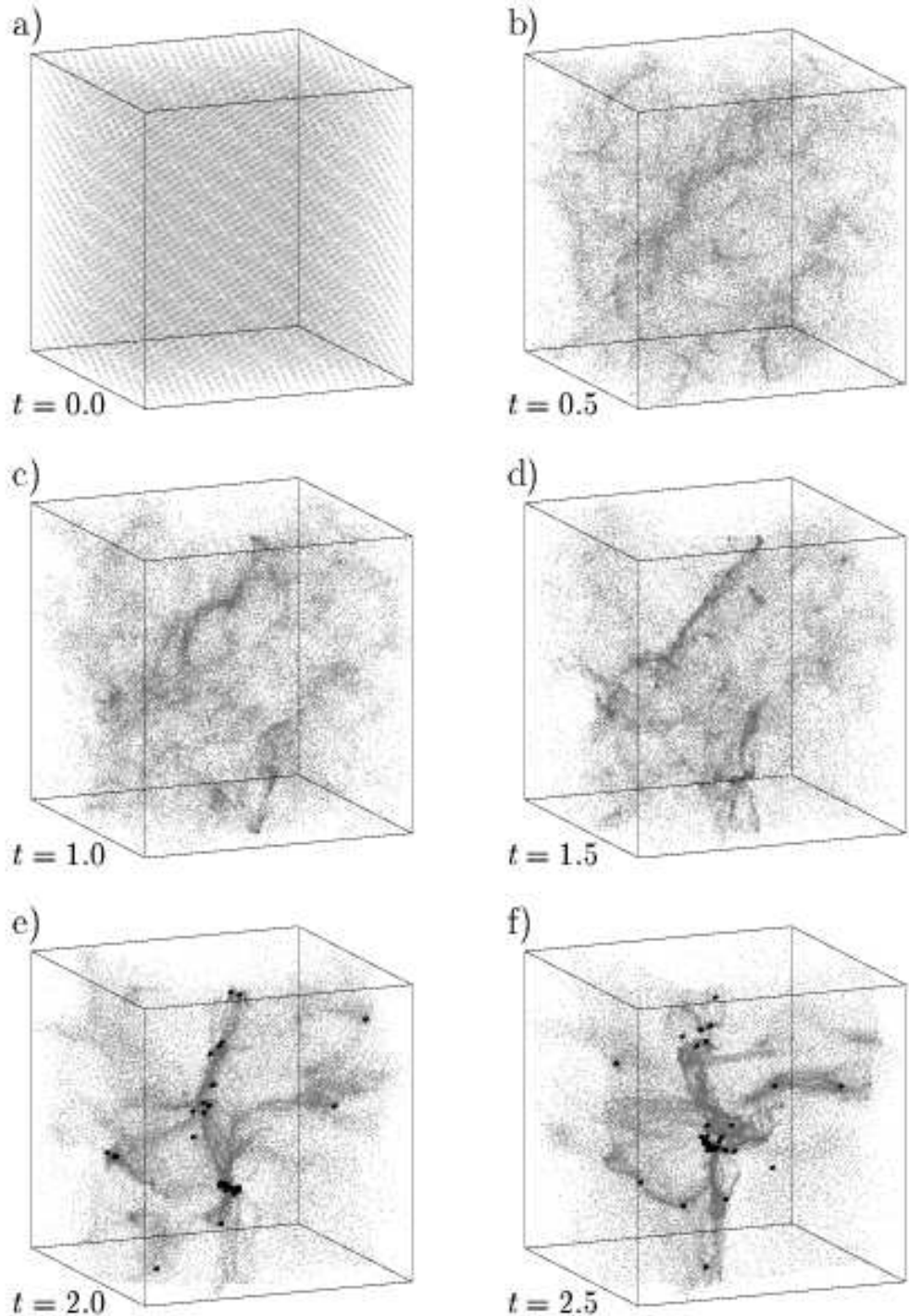}}
\end{picture}
\caption{\label{fig:3D-plot-decay-with-gravity} The 3-dimensional gas
distribution in the SPH simulation of initially supersonic, decaying
turbulence at six different stages of the dynamical evolution. Only
every fourth of the $200\,000$ SPH particles is displayed.  (a) The
first plot shows the homogeneous initial density field. Further
snapshots of the system are taken at (b) $t=0.5$, (c) $t=1.0$, (d)
$t=1.5$, (e) $t=2.0$, and (f) $t=2.5$, where time is measured in units
of the free-fall time scale. The system evolves into a network of
interacting shocks creating a filamentary density structure. As the
turbulent flow decays, local collapse becomes possible. Dense cores
(substituted by `sink' particles) are indicated by dark dots. In (e)
the mass accumulated in collapsed cores is 40\% of the total gas mass,
in (f) this value is 61\%. }
\end{figure}

In this section, I discuss the properties of decaying, initially
supersonic turbulence in a self-gravitating medium. Figure
\ref{fig:3D-plot-decay-with-gravity} displays an SPH simulation with
$200\,000$ particles at six different times of its dynamical
evolution. Since the model is subject to periodic boundary conditions,
every figure has to be considered infinitely replicated in each
direction. Analog to the previous model, the system is initially
homogeneous and its velocity field is generated with $P(k) = {\rm
const.}$ using modes with wave numbers $1\leq k \leq 8$. From the
choice $\alpha = 0.01$ it follows that the system contains 120 {\em
thermal} Jeans masses.  The initial rms velocity dispersion is
$\sigma_v = 0.5$ and with the sound speed $c_{\rm s} = 0.082$ the rms
Mach number follows as $M=6$.  These values imply that the initial
turbulent velocity field contains sufficient energy to globally {\em
stabilize} the system against gravitational collapse.  Scaled to
physical units using a density $n({\rm H_2}) = 10^5\,$cm$^{-3}$, which
is typical for massively star-forming regions (e.g.\ Williams et al.\
2000), the system corresponds to a volume of $[0.32\,{\rm pc}]^3$ and
contains a gas mass of 200$\,$M$_{\odot}$.  As the simulation starts,
the system quickly becomes fully turbulent and loses kinetic energy.
Like in the case without self-gravity a network of intersecting shocks
develops leading to density fluctuations on all scales. If the mass of
a fluctuation exceeds the (local) Jeans limit it begins to contract
due to self-gravity.  During the early evolution, there is enough
kinetic energy to prevent this collapse process on all scales
(Fig.~\ref{fig:3D-plot-decay-with-gravity}b -- $t=0.5$) and the
properties of the system are similar to those of pure hydrodynamic
turbulence. However, as time progresses and turbulent energy decays
the effective Jeans mass decreases.  {\em Local} collapse of shock
generated density fluctuations sets in despite the fact that the
system is still globally stabilized by turbulence (see also Klessen et
al.~2000). The central high-density cores of collapsing clumps are
indicated by black dots. The cores form mainly at the intersection of
filaments, where the density is highest and local collapse is most
likely to set in. When turbulence is decayed sufficiently also
large-scale collapse becomes possible.  Gas clumps follow the global
flow pattern towards a common center of gravity where they may merge
or sub-fragment. Gradually a cluster of dense cores is built up. In
the isothermal model this process continues until all available gas is
accreted onto the `protostellar' cluster (for more details Klessen \&
Burkert 2000).

\begin{figure}[h]
\unitlength1.0cm
\begin{picture}(18,8.0)
\put( -2.25,-16.0){\epsfbox{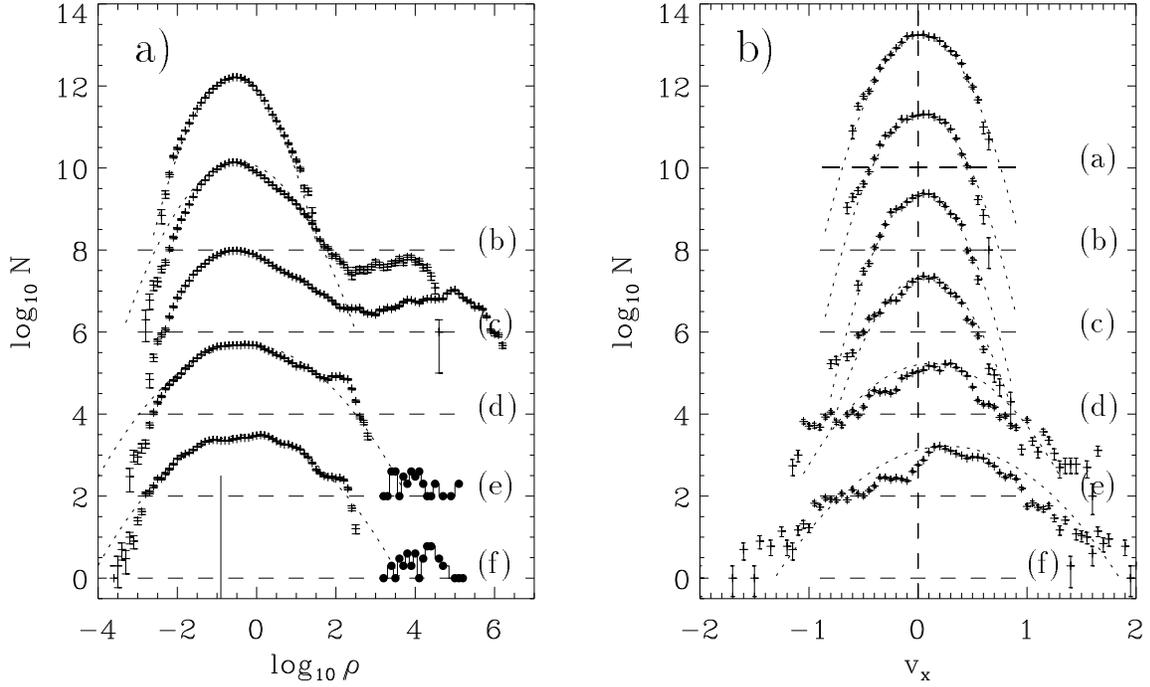}}
\end{picture}
\caption{\label{fig:rho-v-pdf-decay-with-gravity} Pdf's of (a) the
density and (b) the $x$-component of line centroids for the simulation
of initially supersonic, decaying turbulence in self-gravitating
gas. The time sequence is the same as in the previous figure as
denoted by the corresponding letter to the right of each pdf. In the
left panel, the initial density is indicated by the vertical line at
$\rho = 1/8$. The density contributions from collapsed cores forming
in the late stages of the evolution are indicated by solid dots. The
core density corresponds to a mean value computed from the core mass
divided by its accretion volume.  In both figures, each pdf is offset
by $\Delta \log_{10} N = 2.0$ with the base $\log_{10} N = 0.0$
indicated by horizontal dashed lines. The best-fit Gaussian curves are
shown as dotted lines.  }
\end{figure}

\begin{figure}[h]
\unitlength1.0cm
\begin{picture}(18,4.0)
\put( 1.50,-18.1){\epsfbox{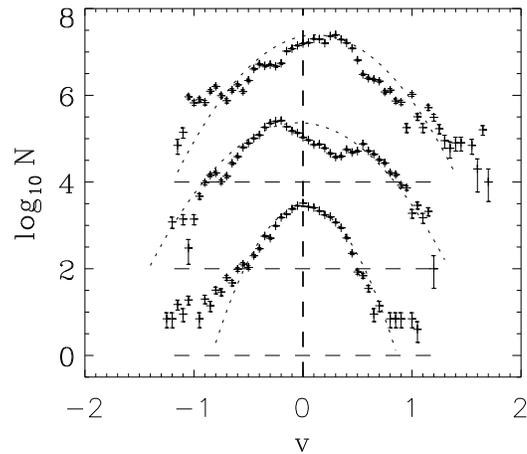}}
\end{picture}
\caption{\label{fig:v-pdf-xyz-decay-with-gravity} Centroid velocity
pdf's for the simulation of initially supersonic, decaying turbulence
in self-gravitating gas at $t=2.0$ for the line-of-sight being along
the $x$-axis (upper curve -- it is identical to the fifth pdf in
Fig.~\ref{fig:rho-v-pdf-decay-with-gravity}b), along the $y$-axis
(middle), and along the $z$-axis of the system (bottom).  Each
distribution is offset by $\Delta \log_{10} N= 2.0$ with the
horizontal lines indicating the base $\log_{10} N= 0.0$. The pdf's of
various projections and velocity components can differ considerably.
}
\end{figure}

The pdf's of (a) the density and of (b) the $x$-component of the line
centroid velocities for the above six model snapshots are displayed in
Fig.~\ref{fig:rho-v-pdf-decay-with-gravity}. The corresponding time is
indicated by the letters at the right side of each panel.  During the
dynamical evolution of the system the density distribution develops a
high density tail. This is the imprint of local collapse. The
densities of compact cores are indicated by solid dots (at $t=2.0$ and
$t=2.5$). Virtually all particles in the high density tails at earlier
times (at $t=1.0$ and more so at $t=1.5$) are accreted onto these
cores. The bulk of matter roughly follows a log-normal density
distribution as indicated by the dotted parabola. The $v$-pdf's are
nearly Gaussian as long as the dynamical state of the system is
dominated by turbulence. Also the width of the pdf remains roughly
constant during this phase. This implies that the decay of turbulent
kinetic energy is in balance with the gain of kinetic energy due to
gravitational (`quasi-static') contraction on large scales. The time
scale for this process is determined by the energy dissipation in
shocks and turbulent eddies.  However, once {\em localized} collapse
is able to set in, accelerations on small scales increase dramatically
and the evolution `speeds up'.  For times $t>2.0$ the centroid pdf's
become wider and exhibit significant deviations from the original
Gaussian shape.  The properties of the pdf's are similar to those
observed in star-forming regions (Miesch \& Scalo 1995, Lis et
al.~1998, Miesch et al.~1998).  This is expected since gravitational
collapse is a necessary ingredient for forming stars.

Gravity creates {\em non}-isotropic density and velocity structure
structures. When analyzing $v$- and $\Delta v$-pdf's, their appearance
and properties will strongly depend on the viewing angle.  This is a
serious point of caution when interpreting observational data, as
molecular cloud are seen only in {\em one} projection.  As
illustration, Fig.~\ref{fig:v-pdf-xyz-decay-with-gravity} plots the
centroid pdf at the time $t=2.0$ for the line-of-sight projection
along all three axes of the system. Whereas the pdf's for the $x$- and
the $y$-component of the velocity centroid are highly structured
(upper and middle curve -- the latter one is even double peaked), the
distribution of the $z$-component (lowest curve) is smooth and much
smaller in width, comparable to the `average' pdf at {\em earlier}
stages of the evolution.  As the variations between different viewing
angles or equivalently different velocity components can be very
large, statements about the 3-dimensional velocity structure from only
observing one projection can be misleading.

\begin{figure}[h]
\unitlength1.0cm
\begin{picture}(18,13.5)
\put( -2.25,-12.75){\epsfbox{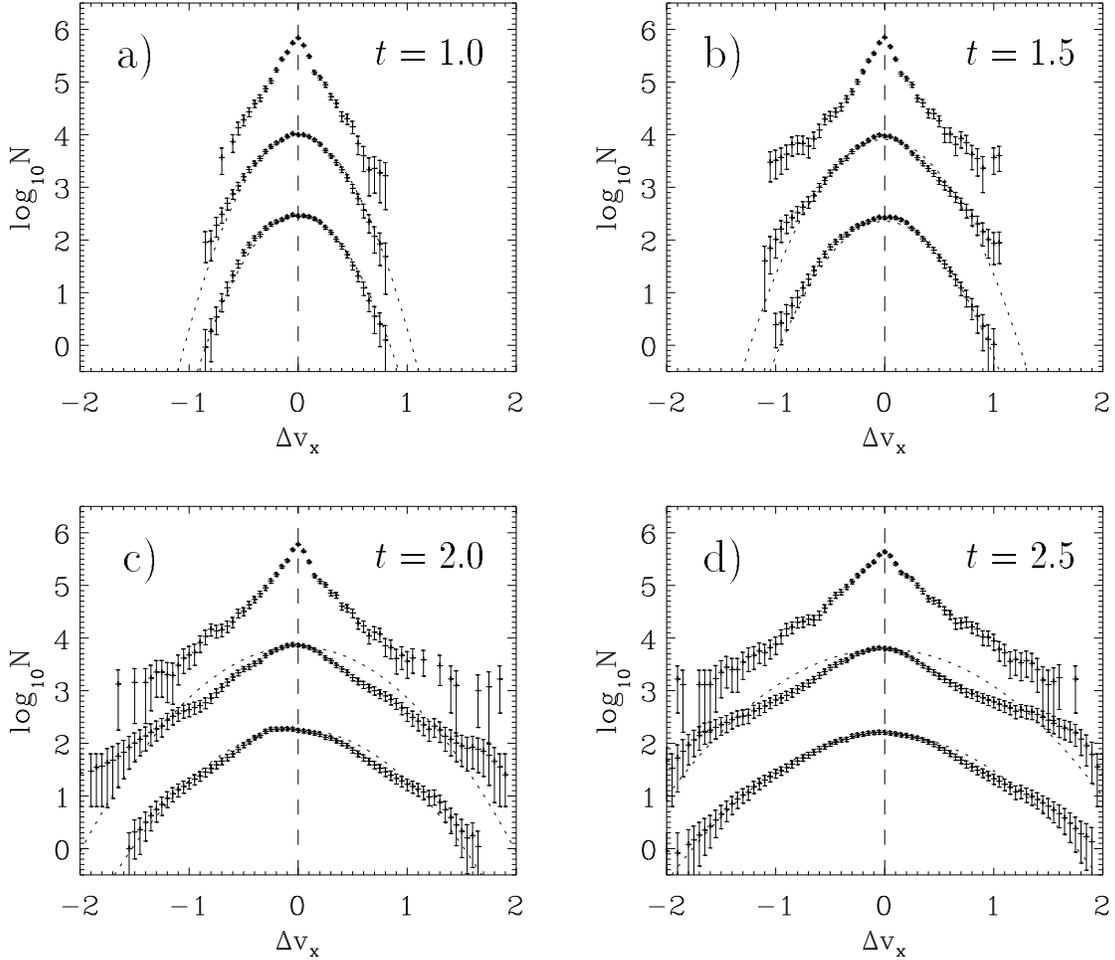}}
\end{picture}
\caption{\label{fig:dv-pdf-decay-with-gravity} Pdf's of the
$x$-component of the centroid velocity increments for three spatial
lags: upper curve -- $\Delta r = 1/32$, middle curve -- $\Delta r =
10/32$, and lower curve -- $\Delta r = 30/32$. The functions are
computed from the simulation of initially supersonic, decaying
turbulence in self-gravitating gas at (a) $t=1.0$, (b) $t=1.5$, (c)
$t=2.0$, and (d) $t=2.5$.  Where appropriate, the Gaussian curves
obtained from the first two moments of the distribution are indicated
by dotted lines. During the early phases of the evolution, the flow is
similar to pure hydrodynamic turbulence (the pdf's are close to the
ones in Fig.~\ref{fig:dv-pdf-decay-no-gravity}). As turbulent energy
decays self-gravity gains influence and the late stages of the
evolution are dominated by gravitational contraction. Consequently the
pdf's in the sequence (a) to (d) become more and more non-Gaussian
with the progression of time. This  concerns the pdf's for
small to intermediate  lags $\Delta r$.  }
\end{figure}

Gravity effects the $\Delta v$-pdf. Figure
\ref{fig:dv-pdf-decay-with-gravity} displays the increment pdf's at
small, intermediate and large spatial lags, analog to Fig.'s
\ref{fig:dv-pdf-init} and \ref{fig:dv-pdf-decay-no-gravity}. Time
ranges from (a) $t=1.0$ to (d) $t=2.5$ corresponding to Fig.'s
\ref{fig:3D-plot-decay-with-gravity}c--f. The pdf's for $t=0.0$ and
$t=0.5$ are not shown since at these stages supersonic turbulence
dominates the dynamic of the system and the pdf's are comparable to
the ones without gravity
(Fig.~\ref{fig:dv-pdf-decay-no-gravity}). This still holds for
$t=1.0$. The increment pdf's for medium to large spatial lags appear
Gaussian, however, the pdf for the smallest lag follows a perfect
exponential all the way inwards to $\Delta v = 0$. Unlike in the case
without gravity, the peak of the distribution is not `round', i.e.~is
not Gaussian in the innermost parts (when scaled to the same
width). {\em It is a sign of self-gravitating systems that the
increment pdf at smallest lags is very strongly peaked and remains
exponential over the entire range of measured velocity
increments}. This behavior is also seen Fig.'s
\ref{fig:dv-pdf-decay-with-gravity}b--d. At these later stages of the
evolution in addition non-Gaussian behavior is also found at medium
lags. This results from the existence of large-scale filaments and
streaming motions. The same behavior is found for the increment pdf's
from observed molecular clouds (for $\rho$-Ophiuchus see Lis et
al~1998; for Orion, Mon R2, L1228, L1551, and HH83 see Miesch et
al.~1998). In each case, the distribution for the smallest lag (one
pixel size) is very strongly peaked at $\Delta v = 0$, in some cases
even more than exponential. The deviations from the Gaussian shape
remain for larger lags but are not so pronounced.  The inclusion of
self-gravity into models of interstellar turbulence leads to good
agreement with the observed increment pdf's. However, this result may
{\em not} be unique as in molecular clouds additional processes are
likely to be present that could also lead to strong deviations from
Gaussianity. 

The time evolution of the statistical moments of the $\Delta v$-pdf's
for various spatial lags is presented in
Fig.~\ref{fig:mom-decay-with-gravity}. It plots (a) the dispersion
$\sigma$, and (b) the kurtosis $\kappa$. The letters on the right-hand
side indicate the corresponding time in
Fig.~\ref{fig:3D-plot-decay-with-gravity}. At $t=0.0$ the width
$\sigma$ of the pdf is approximately constant for all $\Delta r$ and
the kurtosis $\kappa$ is close to normal value of three.  Both
indicate that Gaussian statistics very well describes the initial
velocity field. As turbulent energy decays, gravitational collapse
sets in. Because of the gravitational acceleration, the amplitudes of
centroid velocity differences between separate regions in the cloud
grow larger, the width $\sigma$ of the $\Delta v$-pdf's
increases. This becomes more important when sampling velocity
differences on larger spatial scales, hence $\sigma$ also increases
with $\Delta r$. The slope is $d \log_{10} \sigma/d\log_{10}\Delta r
{\:\lower0.6ex\hbox{$\stackrel{\textstyle <}{\sim}$}\:} 0.2$. For
$\log_{10}\Delta r > -0.4$ it levels out, which is a result of the
adopted periodic boundary conditions. They do not allow for
large-scale velocity gradients. The increasing `peakedness' of $\Delta
v$-pdf is reflected in the large values of the kurtosis $\kappa$ at
the later stages of the evolution. For small spatial lags the pdf's
are more centrally concentrated than exponential (i.e.~$\kappa > 6$),
and even at large spatial separations they are still more strongly
peaked than Gaussian ($\kappa > 3$). The slope at $t=2.5$ is $d
\log_{10} \kappa/d\log_{10}\Delta r \approx -0.4$ which is indeed
comparable to what is found in observed star-forming regions (Miesch
et al.~1998).

For the above simulation of self-gravitating, decaying, supersonic
turbulence, Fig.\ref{fig:dv-array-decay-with-gravity} plots the
2-dimensional distribution of centroid increments for a vector lag
$\Delta \vec{r} = (1/32,1/32)$. The velocity profiles are sampled
along the $x$-axis of the system.  The magnitude of the velocity
increment $\Delta v_{x}$ is indicated at the top of each plot. The
spatial distribution of velocity increments during the initial phases
appears random. Later on, gravity gains influence over the flow and
creates a network of intersecting filaments where gas streams onto and
flows along towards local potential minima. At that stage, the
velocity increments with the highest amplitudes tend to trace the
large-scale filamentary structure. This is the sign of the anisotropic
nature of gravitational collapse motions.

\section{Analysis of Driven Turbulence with Self-Gravity}
\label{sec:driven-with-gravity}
Figure \ref{fig:3D-plot-driven-with-gravity} displays the gas
distribution at different evolutionary stages of a simulation of
driven, supersonic, self-gravitating turbulence. The number of SPH
particles is $205\,379$.  Again, the system is initially homogeneous
in space and has a random Gaussian velocity field with flat power
spectrum in the wave number interval $3 \leq k \leq 4$. It contains 64
{\em thermal} Jeans masses and turbulence is continuously driven as
described in Sec.~\ref{subsec:models}. The initial evolution into
equilibrium between the energy input by the driving force and the
decay of turbulent kinetic energy is computed without self-gravity,
then it is turned on. This phase is displayed in
Fig.~\ref{fig:3D-plot-driven-with-gravity}a. In this state the {\em
turbulent} Jeans mass (on scales larger than the maximum driving wave
length) exceeds the total mass in the system by a factor of two, the
cloud is therefore stabilized by turbulence against gravitational
collapse on global scales. However, local collapse (on scales at or
below the driving scale) is still possible and does occur.  As in the
previous case without driving, the dynamical evolution of the system
leads to the formation of a cluster of dense collapsed cores. This is
shown in Fig.'s \ref{fig:3D-plot-driven-with-gravity}b--d, which
display the system when 20\%, 40\%, and 60\% of the gas mass has
accumulated in dense collapsed cores (at time $t=1.8$, $t=3.2$, and
$t=4.8$, respectively).  However, in the presence of the driving
source the time scales for accretion are longer and the cluster is
less dense.

The pdf's of (a) the density and (b) the $x$-component of the line
centroid velocities corresponding to the above four snapshots are
displayed in Fig.~\ref{fig:rho-v-pdf-driven-with-gravity}.  As in the
previous model, the bulk of gas particles that are not accreted onto
cores build up an approximately log-normal $\rho$-pdf (indicated by
the dotted lines). Also the $v$-pdf remains close to the Gaussian
value. This is different from the case of purely decaying
self-gravitating turbulence, where at some stage global collapse
motions set in and lead to very wide and distorted centroid
pdf's. This is not possible in the simulation of driven turbulence, as
it is stabilized on the largest scales by turbulence. Collapse occurs
only locally which leaves the width of the pdf's relatively unaffected
and only mildly alters their shape.

\begin{figure}[h]
\unitlength1.0cm
\begin{picture}(18,3.5)
\put( -2.25,-15.75){\epsfbox{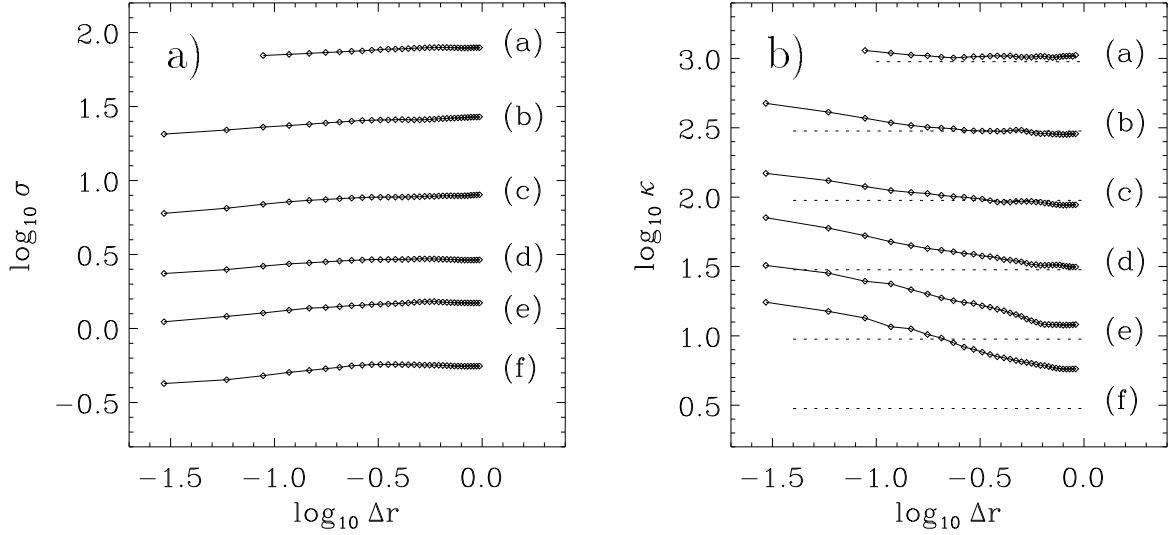}}
\end{picture}
\caption{\label{fig:mom-decay-with-gravity} (a) The second, dispersion
$\sigma$, and (b) the fourth moment, kurtosis $\kappa$, as function of
spatial lag $\Delta r$ for the distribution of velocity increments in
the simulation of self-gravitating, decaying, supersonic
turbulence. The letters on the right-hand side indicate the time at
which the increment pdf's are computed ranging from $t=0.0$ at the top
down to $t=2.5$ at the bottom(see Fig.'s
\ref{fig:3D-plot-decay-with-gravity} or
\ref{fig:rho-v-pdf-decay-with-gravity}). Each pdf is offset by $\Delta
\log_{10} \sigma=0.5$ and $\Delta \log_{10} \kappa = 0.5$,
respectively.}
\end{figure}

\begin{figure}[h]
\unitlength1.0cm
\begin{picture}(18,19.0)
\put( -2.25,-9.0){\epsfbox{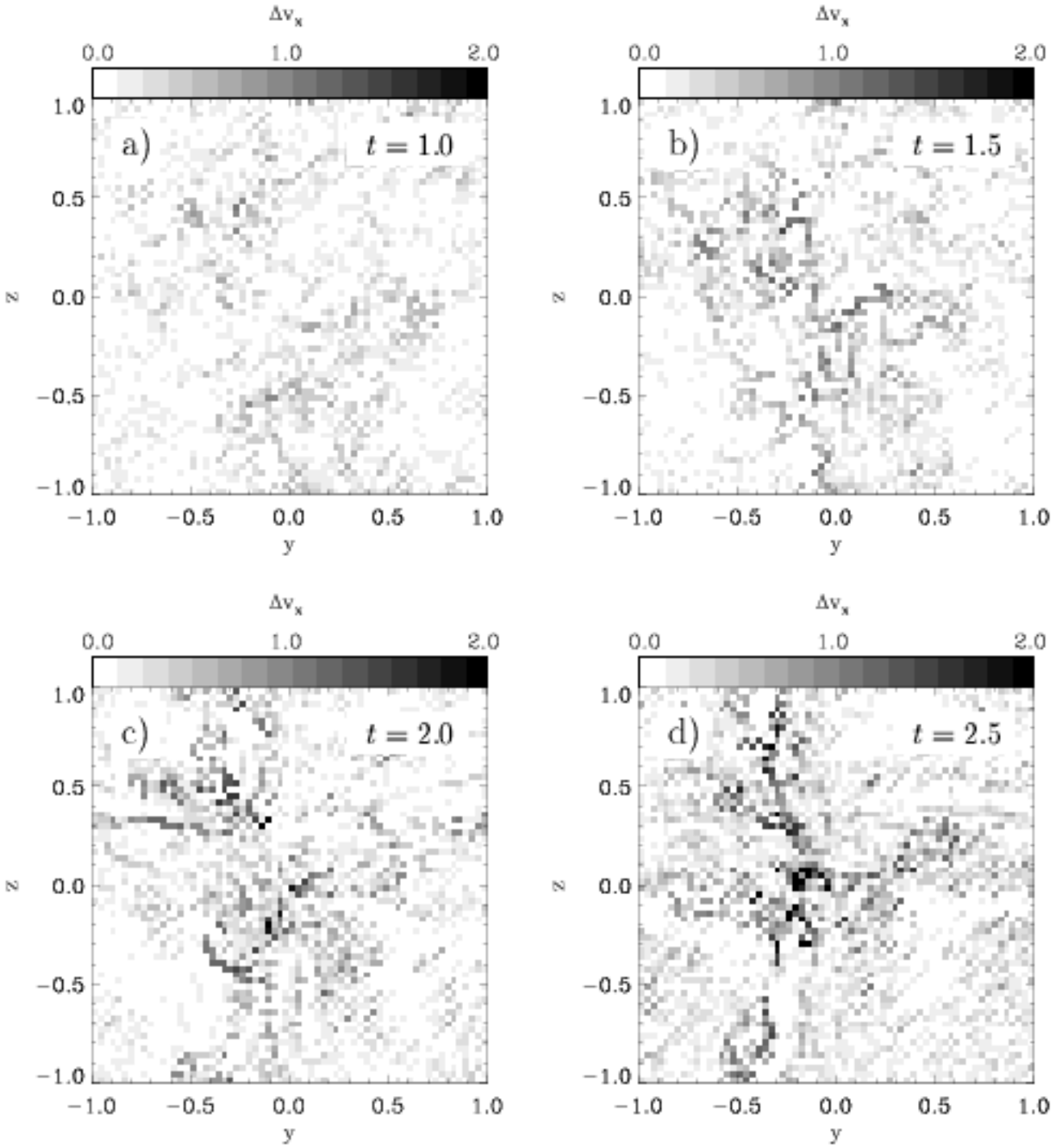}}
\end{picture}
\caption{\label{fig:dv-array-decay-with-gravity} 2-dimensional
distribution (in the $yz$-plane) of centroid increments for velocity
profiles along the $x$-axis of the system between locations separated
by a vector lag $\Delta \vec{r} = (1/32,1/32)$ for the simulation of
self-gravitating, decaying, supersonic turbulence. Analog to
Fig.~\ref{fig:dv-pdf-decay-with-gravity}, the data are displayed for times
(a) $t=1.0$, (b) $t=1.5$, (c) $t=2.0$, and (d) $t=2.5$.  The magnitude
of the velocity increment $\Delta v_{x}$ is indicated at the top of
each plot.  }
\end{figure}

\begin{figure}[h]
\unitlength1.0cm
\begin{picture}(18,12.1)
\put( -0.90,-10.5){\epsfysize=27cm\epsfbox{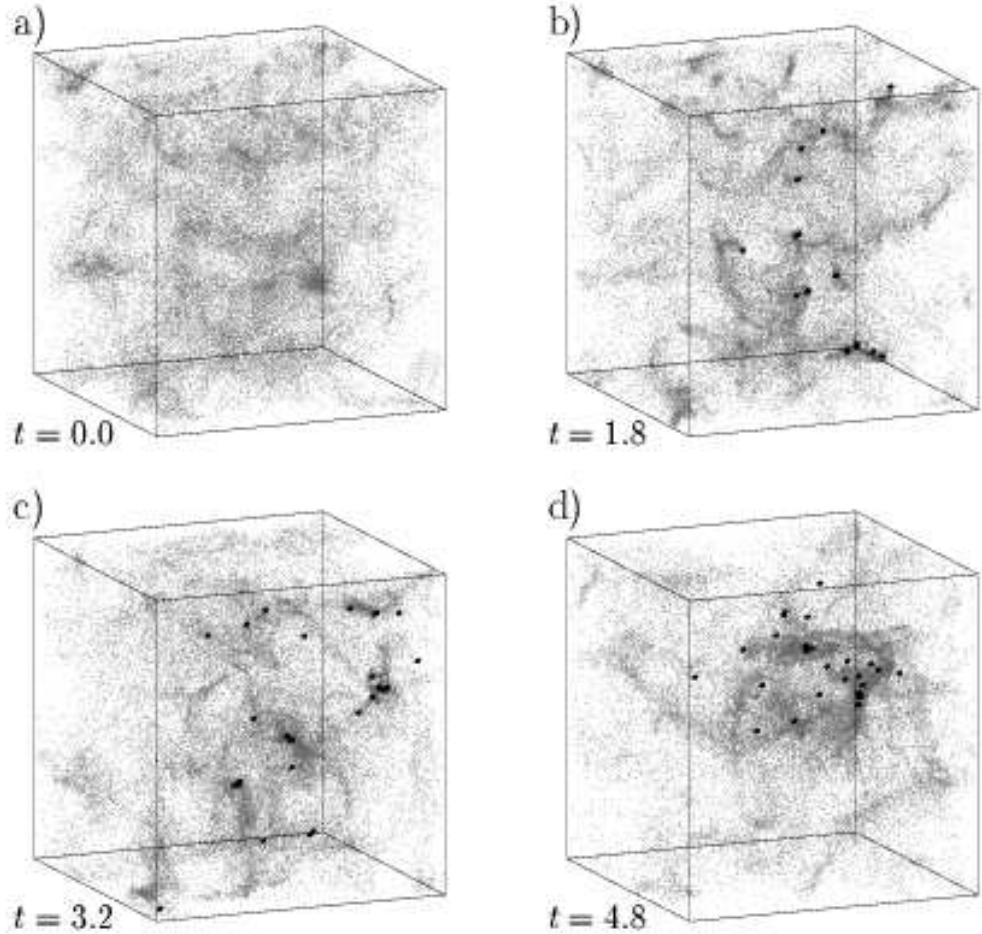}}
\end{picture}
\caption{\label{fig:3D-plot-driven-with-gravity} The 3-dimensional gas
distribution in the simulation of constantly driven turbulence in
self-gravitating gas. Once the turbulent kinetic energy reaches the
equilibrium level, gravity is turned on. This stage is displayed in
(a).  The next three snapshots of the system are taken at times (b)
$t=1.8$, when 20\% of the gas mass is in dense collapsed cores (as
indicated by black dots -- cf. with
Fig.~\ref{fig:3D-plot-decay-with-gravity}), at (c) $t=3.2$, when the
mass in cores is 40\% of the total mass, and at (d) $t=4.8$, when the
cluster of cores contains 60\% of the system mass. Time is given in
units of the free-fall time, but unlike in the previous cases it is
counted from the point gravity is turned on.  }
\end{figure}

\begin{figure}[h]
\unitlength1.0cm
\begin{picture}(18,6.4)
\put( -2.25,-18.0){\epsfbox{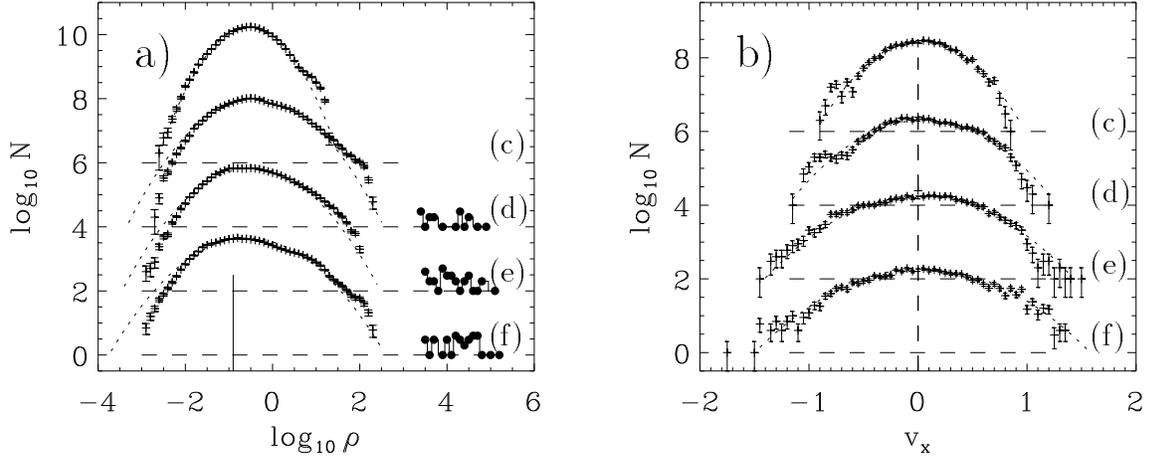}}
\end{picture}
\caption{\label{fig:rho-v-pdf-driven-with-gravity} Pdf's of (a) the
density and (b) the $x$-component of line centroids for the simulation
of driven turbulence in self-gravitating gas. The time sequence is the
same as in the previous figure as indicated by the letters to the
right. Each pdf is offset by $\Delta \log_{10} N = 2.0$ with the base
$\log_{10} N = 0.0$ indicated by horizontal dashed lines. The best-fit
Gaussian curves are shown as dotted lines.  The density contributions
in (a) coming from collapsed cores are indicated by solid dots.  }
\end{figure}

\begin{figure}[h]
\unitlength1.0cm
\begin{picture}(18,6.9)
\put( -2.25,-17.2){\epsfbox{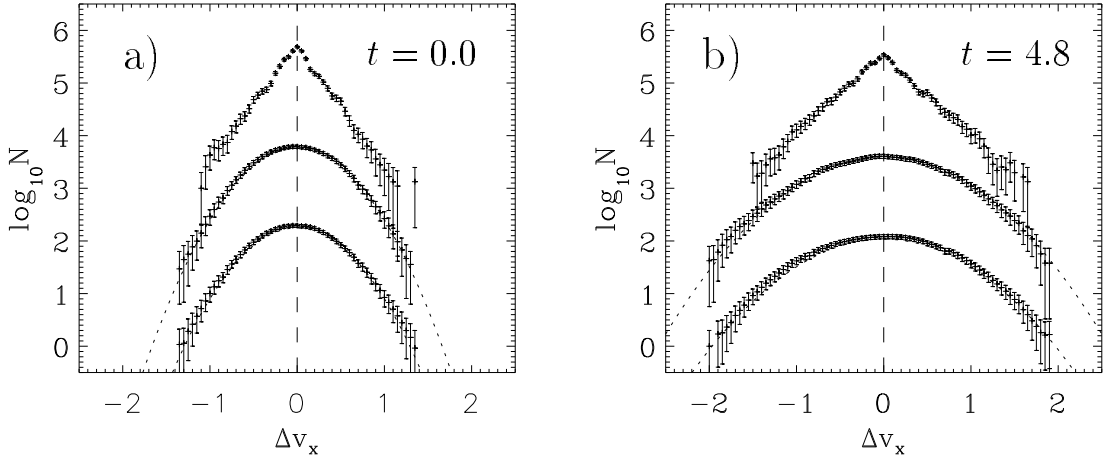}}
\end{picture}
\caption{\label{fig:dv-pdf-driven-with-gravity} Pdf's of the
$x$-component of the centroid velocity increments for three spatial
lags: upper curve -- $\Delta r = 1/32$, middle curve -- $\Delta r =
10/32$, and lower curve -- $\Delta r = 30/32$. The functions are
computed form the simulation of driven, self-gravitating, supersonic
turbulence at (a) $t=0.0$ and (b) $t=4.8$. As in the previous models the increment pdf's
for small spatial lags are approximately exponential, however, the
pdf's for larger separations remain close to Gaussian throughout the
evolution. }
\end{figure}

\begin{figure}[p]
\unitlength1.0cm
\begin{picture}(18,5.0)
\put( -2.25,-16.0){\epsfbox{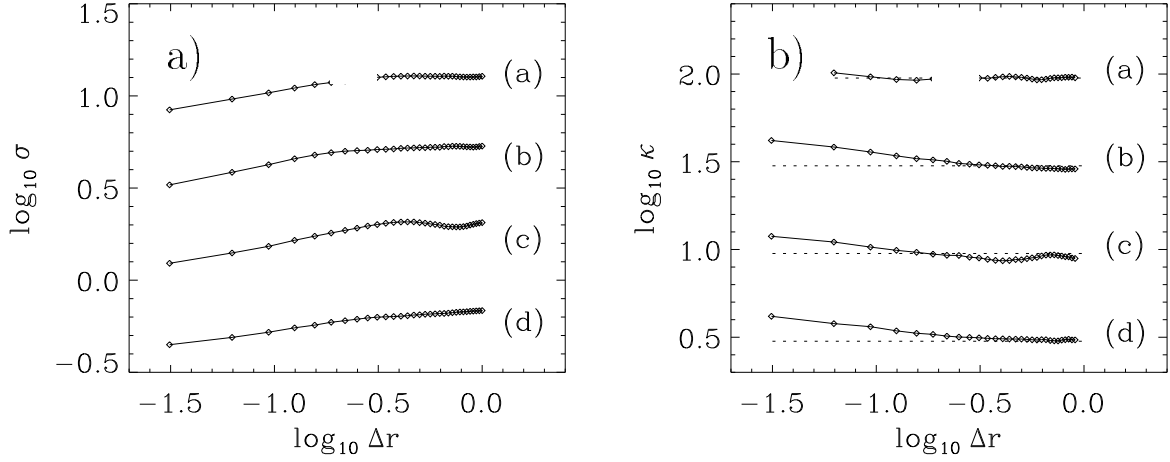}}
\end{picture}
\caption{\label{fig:mom-driven-with-gravity} (a) The second,
dispersion $\sigma$, and (b) the fourth moment, kurtosis $\kappa$, as
function of spatial lag $\Delta r$ for the distribution of velocity
increments in the simulation of driven, self-gravitating, supersonic
turbulence. The letters on the right-hand side indicate again the
correspondence to the times in
Fig.~\ref{fig:3D-plot-driven-with-gravity}.  Each pdf is offset by
$\Delta \log_{10} \sigma=0.5$ and $\Delta \log_{10} \kappa = 0.5$,
respectively.}
\end{figure}

Also the $\Delta v$-pdf's show no obvious sign of evolution. For the
$x$-component of the velocity these functions are displayed in
Fig.~\ref{fig:dv-pdf-driven-with-gravity}, again for three different
spatial lags. The chosen times correspond (a) to the equilibrium state
at $t=0.0$, and (b) to $t=4.8$ which is the final state of the
simulation. The pdf's only marginally grow in width. At every
evolutionary stage, the pdf for the smallest spatial lag is
exponential, whereas the pdf's for medium and large shift vectors
closely follow the Gaussian curve defined by the first two moments of
the distribution (dotted lines). The functions are similar to the ones
in the previous model before the large scale collapse motions set in
(Fig.~\ref{fig:dv-pdf-decay-with-gravity}a, b). Only overall
contraction will affect $\Delta v$-pdf at medium to large lags. This
behavior also follows from comparing the statistical moments. Figure
\ref{fig:mom-driven-with-gravity} plots (a) the dispersion $\sigma$
and (b) the kurtosis $\kappa$ as function of the spatial lag $\Delta
r$.  Figures \ref{fig:mom-decay-with-gravity}a and
\ref{fig:mom-driven-with-gravity}a are very similar, as soon as
turbulence is established the width $\sigma$ of the pdf increases with
$\Delta r$ with a slope of $d \log_{10} \sigma/d\log_{10}\Delta r
{\:\lower0.6ex\hbox{$\stackrel{\textstyle <}{\sim}$}\:} 0.2$ for small
to medium lags and levels out for larger ones. However, when comparing
the `peakedness' of the pdf as indicated by $\kappa$ (Fig.'s
\ref{fig:mom-decay-with-gravity}b and
\ref{fig:mom-driven-with-gravity}b) the model of decaying
self-gravitating turbulence yields much higher values since the pdf's
are more strongly peaked due to the presence of large-scale collapse
motions.

\begin{figure}[p]
\unitlength1.0cm
\begin{picture}(18,10.0)
\put( -2.25,-16.0){\epsfbox{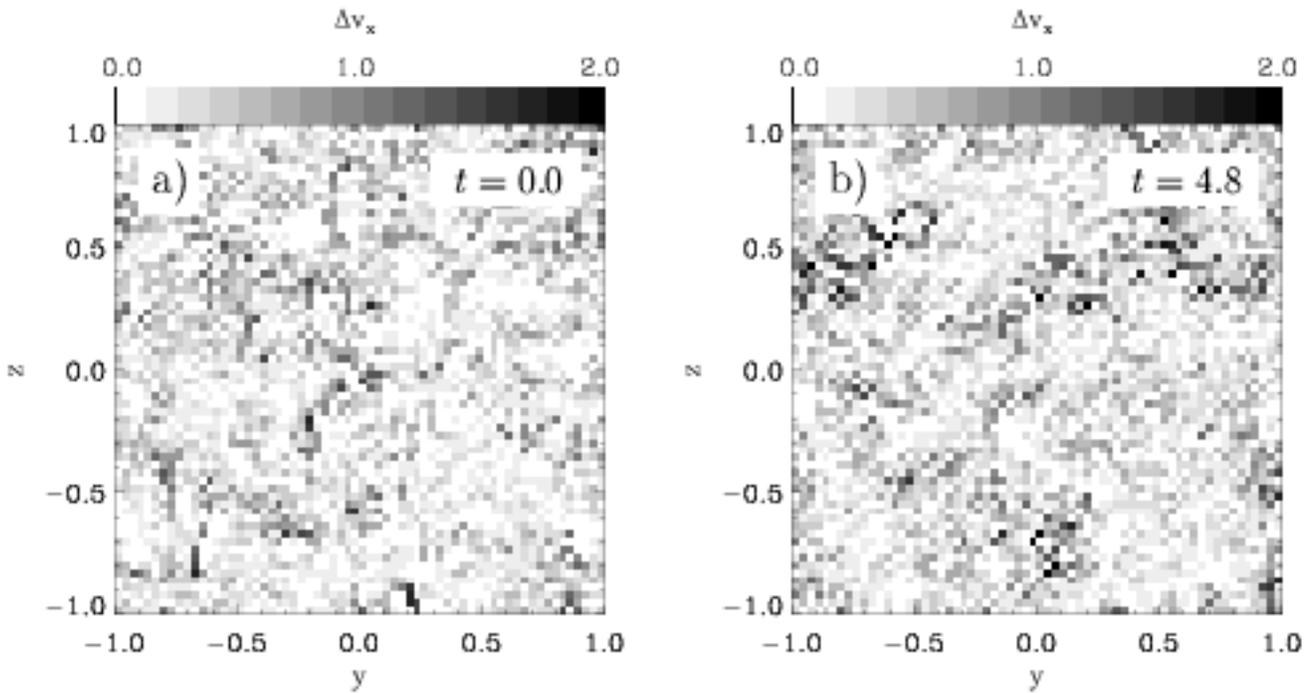}}
\end{picture}
\caption{\label{fig:dv-array-driven-with-gravity} 2-dimensional
distribution (in the $yz$-plane) of the absolute value of the
$x$-component of centroid velocity increments between locations
separated by a vector lag $\Delta \vec{r} = (1/32,1/32)$ for the
simulation of driven self-gravitating supersonic turbulence. The data
are displayed at times (a) $t=0.0$, and (b) $t=4.8$.  The scaling is
indicated at the top of each figure.  }
\end{figure}

Figure \ref{fig:dv-array-driven-with-gravity} finally shows the
spatial distribution of the $x$-component of the line centroid
increments for a vector lag $\Delta \vec{r} = (1/32,1/31)$. Since the
increment maps at different evolutionary times are statistically
indistinguishable, only times (a) $t=0.0$ and (b) $t=4.8$ are
displayed in the figure. As in the case of supersonic, purely
hydrodynamic turbulence the spatial distribution of velocity
increments appears  random and uncorrelated.

The adopted driving mechanism prevents global collapse. The bulk
properties of the system therefore resemble hydrodynamic, {\em
non}-self-gravitating turbulence. However, local collapse motions do
exist and are responsible for noticeable distortions away from the
Gaussian statistics. As the non-local driving scheme adopted here
introduces a bias towards Gaussian velocity fields, these distortions
are not very large. There is a need to introduce other, more realistic
driving agents into this analysis. These could lead to much stronger
non-Gaussian signatures in the pdf's.

\section{Summary}
\label{sec:summary}

SPH simulations of driven and decaying, supersonic, turbulent flows
with and without self-gravity have been analyzed in this study . It
extends previous investigations of mildly supersonic, decaying, {\em
non}-self-gravitating turbulence (Lis et al.~1996, 1998) into a regime
more relevant molecular clouds, by (a) considering highly supersonic
flows and by including (b) self-gravity and (c) a driving source for
turbulence.

The flow properties are characterized by using the probability
distribution functions of the density, of the line-of-sight velocity
centroids, and of their increments. Furthermore the dispersion and the
kurtosis of the increment pdf's are discussed, as well as the spatial
distribution of the velocity increments for the smallest spatial lags.

(1) To asses the influence of variance effects, simple Gaussian
velocity fluctuations are studied. The insufficient sampling of random
Gaussian ensembles leads to distorted pdf's similar to the observed
ones. For line profiles this has been shown by Dubinski et al.~(1995).

(2) Decaying, initially highly supersonic turbulence without
self-gravity leads to pdf's which also exhibit deviations from
Gaussianity. For the trans- and subsonic regime this has been reported
by Lis et al.~(1996, 1998). However, neglecting gravity and thus not
allowing for the occurance of collapse motions, these distortions are
not very pronounced and cannot account well for the observational data
(Lis et al.~1998, Miesch et al.~1998).

(3) When including gravity into the models of decaying initially
supersonic turbulence, the pdf's get into better agreement with the
observations. During the early dynamical evolution of the system
turbulence carries enough kinetic energy to prevent collapse on all
scales. In this phase the properties of the system are similar to
those of non-gravitating hydrodynamic supersonic turbulence. However, as
turbulent energy decays gravitational collapse sets in. First
localized and on small scales, but as the turbulent support continues
to diminish collapse motions include increasingly larger spatial
scales. The evolution leads to the formation of an embedded cluster of
dense protostellar cores (see also Klessen \& Burkert 2000). As the
collapse scale grows, the $\rho$-, $v$-, and $\Delta v$-pdf's get
increasingly distorted. In particular, the $\Delta v$-pdf's for small
spatial lags are strongly peaked and exponential over the entire range
of measured velocities. This is very similar to what is observed in
molecular clouds (for $\rho$-Ophiuchus see Lis et al~1998; for Orion,
Mon R2, L1228, L1551, and HH83 see Miesch et al.~1998).

(4) The most realistic model for interstellar turbulence considered
here includes a simple (non-local) driving scheme. It is used to
stabilize the system against collapse on large scales. Again
non-Gaussian pdf's are observed. Despite global stability, local
collapse is possible and the system again evolves towards the
formation of an embedded cluster of accreting protostellar cores.  As
the adopted driving scheme introduces a bias towards maintaining a
Gaussian velocity distribution, the properties of the pdf's fall in
between the ones of pure hydrodynamic supersonic turbulence and the
ones observed in systems where self-gravity dominates after sufficient
turbulent decay.  This situation may change when considering more
realistic driving schemes.

(5) A point of caution: The use of $v$- and $\Delta v$-pdf's to
unambiguously characterize interstellar turbulence and to identify
possible physical driving mechanisms may be limited.  {\em All} models
considered in the current analysis lead to non-Gaussian signatures in
the pdf's, differences are only gradual. In molecular clouds the
number of physical processes that are expected to give rise to
deviations from Gaussian statistics is large. Simple statistical
sampling effects (Sec.~\ref{sec:init}) and turbulent intermittency
caused by vortex motion (Lis et al.~1996, 1998), as well as the effect
self-gravity (Sec.~\ref{sec:decay-with-gravity}) and of shock
interaction in highly supersonic flows (Mac~Low \& Ossenkopf 2000),
{\em all} will lead to non-Gaussian signatures in the observed pdf's.
Also stellar feedback processes, galactic shear and the presence of
magnetic fields will influence the interstellar medium and create
distortions in the velocity field. This needs to be studied in further
detail.  In addition, the full 3-dimensional spatial and kinematical
information is not accessible in molecular clouds, measured quantities
are always projections along the line-of-sight. As the structure of
molecular clouds is extremely complex, the properties of the pdf's may
vary considerably with the viewing angle. Attempts to disentangle the
different physical processes influencing interstellar turbulence
therefore should no rely on analyzing velocity pdf's alone, they
require additional statistical information.

\acknowledgements I thank A.~Burkert, F.~Heitsch, and M.-M.~Mac~Low
for many fruitful and stimulating discussions, and the editor S.~Shore
for his comments on the paper and his help with an extremely slow and
non-responsive (anonymous) referee.


\clearpage

\end{document}